\documentclass[a4paper,fleqn,usenatbib]{mnras}

\usepackage{newtxtext,newtxmath}

\usepackage[T1]{fontenc}
\usepackage{ae,aecompl}

\usepackage{graphicx}	
\usepackage{amsmath}	
\usepackage{amssymb}	
\usepackage[export]{adjustbox}
\usepackage{natbib}
\usepackage{gensymb}
\usepackage{tikz}
\usepackage{amsmath}
\usepackage{multirow}
\usetikzlibrary{decorations.markings}

\DeclareMathAlphabet{\mathsc}{OT1}{cmr}{m}{sc}
\def\testbx{bx}%
\DeclareRobustCommand{\ion}[2]{%
	\relax\ifmmode
	\ifx\testbx\f@series
	{\mathbf{#1\,\mathsc{#2}}}\else
	{\mathrm{#1\,\mathsc{#2}}}\fi
	\else\textup{#1\,{\mdseries\textsc{#2}}}%
	\fi}

\newcommand{\HI}{\ion{H}{i}}
\newcommand{\HIII}{\ion{H}{}}
\newcommand{\HII}{\ion{H}{ii}}

\newcommand{\CIV}{\ion{C}{iv}}

\newcommand{\FeII}{\ion{Fe}{ii}}
\newcommand{\FeN}{\ion{Fe}{}}
\newcommand{\FeIII}{\ion{Fe}{iii}}

\newcommand{\MgII}{\ion{Mg}{ii}}

\newcommand{\OII}{\ion{O}{ii}}
\newcommand{\OIII}{\ion{O}{iii}}

\newcommand{\kms}{km\,s$^{-1}$}
\newcommand{\zabs}{$z_{\rm abs}$}
\newcommand{\msun}{M$_\odot$}
\newcommand{\nhi}{$N$(\HI)}
\title[Galactic wind using MUSE]{A Lyman limit system associated with galactic winds \thanks{Based on data obtained under the ESO programmes 096.A-0303 and 095.A-0338 at the European Southern Observatories with MUSE at the 8.2 m telescopes operated at the Paranal Observatory, Chile.}}

\author[Hadi Rahmani et al.]{Hadi Rahmani,$^{1,2}$\thanks{E-mail: hadi.rahmani@gmail.com}
C\'{e}line P\'{e}roux,$^{2}$
Ilane Schroetter,$^{1}$
Ramona Augustin,$^{2,3}$
\newauthor Nicolas Bouch\'e,$^4$
Jens-Kristian Krogager,$^{5}$
Varsha P. Kulkarni,$^{6}$
Bruno Milliard,$^{2}$
\newauthor Palle M\o{}ller,$^{3}$
Max Pettini,$^{7}$
Jo\"{e}l Vernet,$^{3}$
and Donald G. York$^{8,9}$
\\
$^1$GEPI, Observatoire de Paris, PSL Universit\'{e}, CNRS,  5 Place Jules Janssen, 92190 Meudon, France\\
$^2$Aix Marseille Universit\'{e}, CNRS, LAM (Laboratoire d'Astrophysique de Marseille) UMR 7326, 13388, Marseille, France\\
$^{3}$European Southern Observatory, Karl-Schwarzschildstrasse 2, D-85748 Garching bei M{\"u}nchen, Germany\\
$^{4}$CNRS/IRAP, 9 Avenue Colonel Roche, F-31400 Toulouse, France\\
$^{5}$Institut d'Astrophysique de Paris, CNRS-UPMC, UMR7095, 98bis bd Arago, 75014 Paris, France\\
$^{6}$Department of Physics and Astronomy, University of South Carolina, Columbia, SC 29208, USA\\
$^{7}$Institute of Astronomy, University of Cambridge, Madingley Road, Cambridge CB3 0HA, UK\\
$^{8}$Department of Astronomy and Astrophysics, The University of Chicago, Chicago, IL 60637, USA\\
$^{9}$Enrico Fermi Institute, The University of Chicago, Chicago, IL 60637, USA
}

\pubyear{2017}

\begin{document}
\label{firstpage}
\pagerange{\pageref{firstpage}--\pageref{lastpage}}
\maketitle

\begin{abstract}
Projected quasar galaxy pairs provide powerful means to study the circumgalactic medium (CGM) that maintains the relics of galactic feedback and the accreted gas from the intergalactic medium. Here, we study the nature of a Lyman Limit system (LLS) with N(\HI)=10$^{19.1\pm0.3}$\,cm$^{-2}$ and a dust-uncorrected metallicity of [Fe/H]$=-1.1\pm0.3$ at $z=0.78$ towards Q0152$-020$. The \MgII\ absorption profiles are composed of a main saturated and a few weaker optically thin components. Using MUSE observations we detect one galaxy close to the absorption redshift at an impact parameter of $54$\,kpc. This galaxy exhibits nebular emission lines from which we measure a dust-corrected star formation rate of $10^{+8}_{-5}$\,M$_\odot$\,yr$^{-1}$ and an emission metallicity of [O/H]$=-0.1\pm0.2$. 
By combining the absorption line kinematics with the host galaxy morphokinematics we find that while the main absorption component can originate from a galactic wind at $V_{\rm w}=110\pm4$\,\kms\ the weaker components cannot.
We estimate a mass ejection rate of $\dot M\gtrsim0.8$\,M$_\odot$\,yr$^{-1}$ that translates to a loading factor of $\eta\gtrsim0.1$. Since the local escape velocity of the halo, $V_{\rm esc}\simeq430$\,\kms, is a few times larger than $V_{\rm w}$, we expect this gas will remain bound to the host galaxy. These observations provide additional constraints on the physical properties of winds predicted by galaxy formation models. We also present the VLT/X-Shooter data analysis of 4 other absorbing systems at $1.1<z<1.5$ in this sightline with their host galaxies identified in the MUSE data.
\end{abstract}

%
%
%
\section{Introduction}
The circumgalactic medium (CGM) is the region around a galaxy through which the baryon exchange between the intergalactic medium (IGM) and the galaxy occurs. On the one hand it hosts the fresh pristine gas entering the halo from the IGM but on the other hand it contains relics of metal enriched galactic winds. While the extent of the CGM varies based on the definition, usually it is considered as the region with a radius of $\sim300$\,kpc surrounding galaxies \citep[e.g.,][]{Steidel10,Shull14_a}. To understand the full cycle of gas from IGM into galaxies and from galaxies towards IGM it is crucial to understand the distribution, kinematics and metal content of the gas in the CGM.

Due to the very low density of the gas it is not yet possible to probe the CGM in emission \citep[but see,][]{Frank12}. Absorption lines towards background quasars provide the most sensitive tools to carry out CGM studies. In particular, quasar absorbers with large \HI\ column density, N(\HI)$\gtrsim10^{19}$\,${\rm cm^{-2}}$, afford robust measurements of the metal content of the CGM. There are several studies that also show that the majority of the neutral gas reservoirs in the Universe are associated with such absorbers \citep{Peroux03,Prochaska05,Noterdaeme09dla,Noterdaeme12dla,Zafar13}.  Therefore, by studying such strong \HI\ absorbers one not only probes the CGM but also traces the possible fuel for forming stars in galaxies.

\citet{Quiret16} reported a bimodal distribution of metallicity for a subsample of strong \HI\ absorbers ($19.0<\log [N(\HI)/{\rm cm}^{-2}] <20.3$) at $z<1$. The high metallicity side of the distribution could be interpreted as a possible tracer for galactic winds and the low metallicity side as a possible tracer for the pristine IGM gas accretion \citep[see also][for similar results but for Lyman limit systems (LLS) with N(\HI)$>10^{16}$\,cm$^{-2}$]{Lehner13,Wotta16}. \citet{Hafen16} studied a sample of LLS at similar redshifts extracted from hydrodynamical simulations where they demonstrated that such absorbers are associated with gaseous complexes in the CGM. While their mean LLS metallicity ([X/H]=$-0.9$) is consistent with that of observations, they did not find a bimodal metallicity distribution. It is worth also noting that in their simulation, \citet{Hafen16} did not find as many low metallicity systems as those detected in observations. However, to disentangle the role of the CGM gas one needs to study together the LLS and its host galaxy. 

The CGM can also be studied from the metal absorption lines in the spectra of the host galaxies themselves, the so called down-the-barrel technique \citep[e.g.,][]{Pettini02_dtb,Steidel10,Rubin12,Martin12,Kacprzak14,Erb15,Bordoloi16,Finley17}. Blueshifted and redshifted absorption lines are produced in the spectra of galaxies in the cases of, respectively, galactic winds and infalling gas. Large sample studies have been successful in finding predominantly outflows; infalling gas has been unambiguously identified in only a few cases \citep{Rubin12,Martin12}. \citet{Rubin12} postulated that the strong and broad absorption lines from outflowing gas can smear out the absorption signature from infalling gas. In any case, the low resolution spectra of galaxies and the unknown distance between absorbing gas and the host galaxy limit the ultimate accuracy of the derived physical parameters from down-the-barrel studies.

The nature of the gas in the CGM can be studied in more detail when the host galaxies of quasar absorbers are also detected in emission \citep{Moller98a,Rahmani10,Peroux11a,Peroux11b,Noterdaeme12,Bouche13,Christensen14,Rahmani16,Rahmani18,Moller17}. As an example \citet{Noterdaeme12} studied multiple emission lines from the host galaxy of a strong \HI\ absorber at $z=2.2$ and demonstrated that the absorption is most likely associated with an outflow \citep[see also,][]{Kulkarni12}. \citet{Krogager12} observed a tight anti-correlation between the impact parameter and the N(\HI) which is likely driven by feedback mechanisms \citep[see also][for similar results]{Monier09,Rao11,Rahmani16}.
 
Much better understanding on the nature of the CGM can be achieved if one combines the absorption line kinematics with those of the host galaxy absorber extracted from integral field unit (IFU) observation. As part of the SIMPLE survey \citep{Bouche07_simple,Bouche12a}, in a study of the host galaxies of strong \MgII\ absorbers, \citet{Schroetter15} demonstrated that such absorbers are often associated with galactic winds. \citet{Peroux11a} used VLT/SINFONI IFU observations of the field of a handful of Damped Lyman-$\alpha$ systems (that are absorbers having $\log$\,[$N(\HI)$/cm$^{-2}$]$>20.3$) at $z\sim1$ to study the kinematics of their host galaxies. They also showed that the detection rate is higher for the subsample of absorbers with higher metallicity. \citet{Rahmani18} used VLT/MUSE observation of an LLS host galaxy along with high resolution Keck/HIRES data of the quasar absorption line to show that the LLS originates from a warped disk. Such warped disks are probably associated with cold gas accreting onto galaxies \citep[see also][for similar studies]{Bouche13,Burchett13,Diamond-Stanic16}.

Here we present the detailed study of an LLS at $z=0.78$ towards Q0152$-020$\footnote{RA(J2000)=28.113464, Dec(J2000)=$-20.018435$.} \citep{Sargent88_MgII,Rao06}. We combine new VLT/MUSE observation of the quasar field with high spectral resolution Keck/HIRES, low spectral resolution HST/FOS spectroscopy of the quasar and HST/WFPC2 $F702W$ high spatial resolution image to understand the nature of the CGM gas absorber. The paper is organized as follows. In section \ref{section_observation} we provide a summary of the available observations. In section \ref{section_absline} we study the absorption line and in section \ref{section_host_gal} we present the properties of the host galaxy absorber. In section \ref{section_conclude} we study the nature of the gas we see in absorption and finally we summarize the paper in section \ref{section_summary}. We also mention other absorbing systems in this sightline having redshifts $1.0<z<1.5$ in appendices \ref{appendix_absline} and \ref{appendix_hostgal}. Throughout this paper, we assume a flat $\Lambda$CDM cosmology with $H_0=69.3$\,\kms\,Mpc$^{-1}$ and $\Omega_m=0.286$ \citep{Hinshaw13}.
\section{Observations of the field of Q0152$-$020}\label{section_observation}
In this section we summarize the spectroscopic and imaging observations of the field of Q0152$-$020. Apart from VLT/X-Shooter observations the remaining data have been discussed in detail by \citet{Rahmani18}. Hence, while we present the X-Shooter observations in the appendix \ref{XSH_OBS}, we only summarize the rest of the data and refer the reader to \citet{Rahmani18} for a comprehensive description.

A field of view of $\sim2$\,arcmin$^2$ around this quasar has been observed with Hubble Space Telescope Wide-Field Planetary Camera 2 (HST/WFPC2) $F702W$ with a $0.1''$ pixel scale and resolution $FWHM=0.2''$ (PI: Steidel, Program ID: 6557). The available Keck/HIRES spectrum of this quasar \citep{OMeara15} covers a wavelength range of 3245\,\AA\ to 5995\,\AA\ and has a spectral resolution of $R=48000$. IFU data of Q0152$-$020 has been obtained from VLT/MUSE observations (program 96.A-0303, PI: P\'{e}roux) with 100 minutes exposure time. We refer the reader to \citet{Peroux17} for a complete description of the procedure used to reduce these VLT/MUSE data. The VLT/MUSE data have a spectral resolution of 2.6\,\AA\ with a sampling of 1.25 \AA/pixel and a wavelength coverage of 4750\,\AA--9350\,\AA. Our final reduced cube has its wavelength scale in air and hence to obtain the redshifts we use the rest-frame wavelengths of nebular transitions as measured in air. The results remain the same if we convert the wavelength scale of the cube into vacuum and utilize the rest-frame wavelengths in vacuum.

It has been demonstrated that the light distribution of point sources in VLT/MUSE can be well described by a Moffat function \citep{Bacon15,Contini16}. To obtain the seeing of the MUSE final combined data we generated narrow band images of the quasar over several wavelength ranges and modeled them using \textsc{galfit} \citep{Peng10AJ}. As a typical value we found a $FWHM=0.6''$ ($\equiv2.95$ pixel) at $\lambda\sim6640$\,\AA. We neglected the spatial variations of the $FWHM$ over the MUSE field of view as it has been shown to be only a few percent \citep{Contini16}.

To extract the spectra of different objects from the MUSE cube we use the MUSE \textsc{Python} Data Analysis Framework  (\textsc{mpdaf v2.0}) which is a Python based package, developed by the MUSE Consortium \citep{Piqueras17}.
\begin{figure*}
	\centering	
	\includegraphics[width=1.\hsize,bb=22 211 588 579,clip=,angle=0]{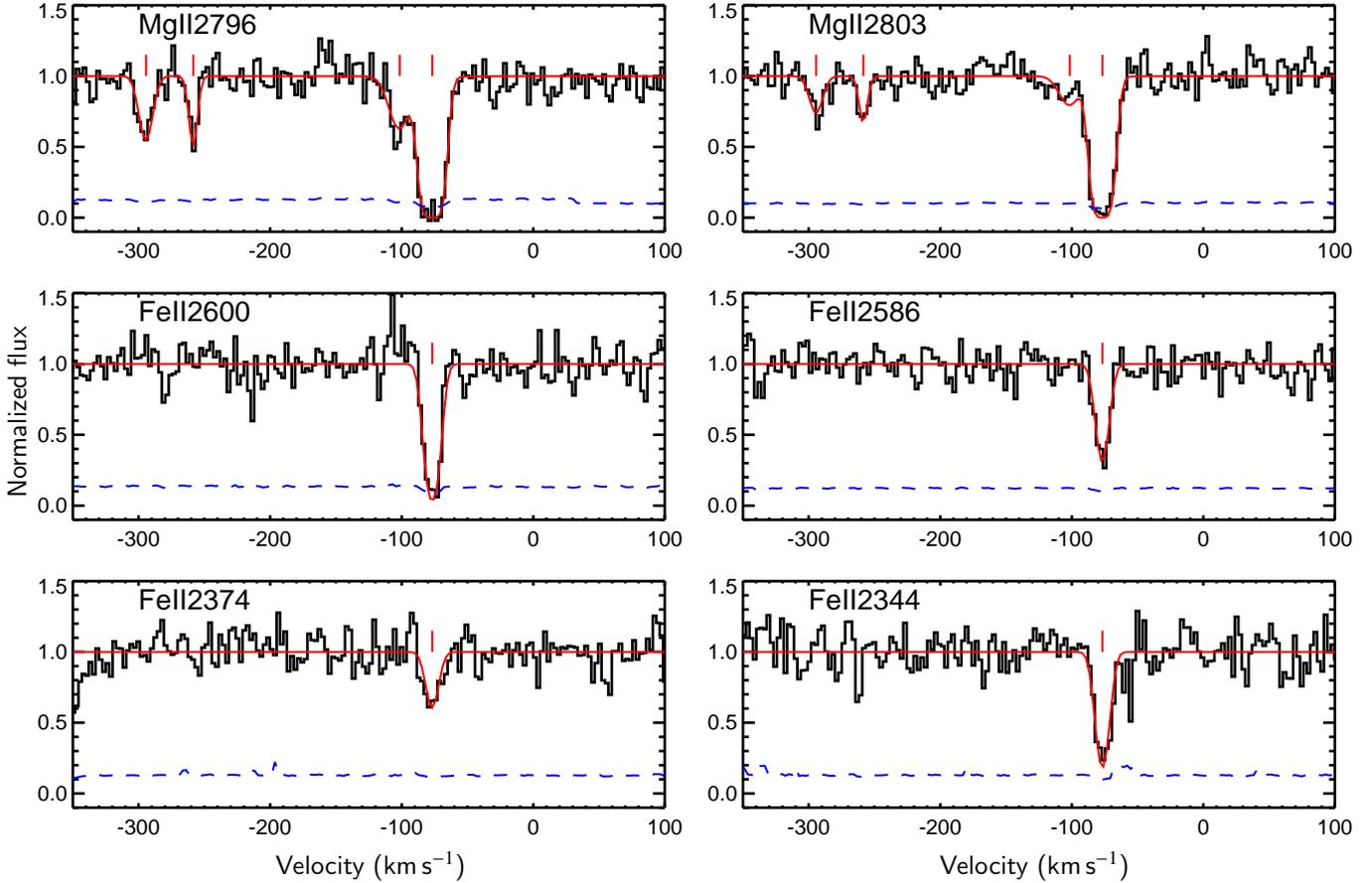} 	
		\vspace*{.05cm}
	\caption{Normalized Keck/HIRES quasar absorption line spectrum. The zero velocity is set at the systemic redshift of the detected galaxy, $z=0.78025$. Black histogram and red continuous lines represent the observed and modeled spectra respectively. The blue dashed lines show the one sigma photon noise at each datapoint. The position of Voigt profile components are indicated using vertical tick marks. }
	\begin{picture}(0,0)(0,0)
	\put(-263,190){ \rotatebox{90}{ \large \sffamily Normalized flux}} 
		\put(-165,50){ \rotatebox{0}{ \large \sffamily Velocity (km\,s$^{-1}$)}} 
		\put(85,50){ \rotatebox{0}{ \large \sffamily Velocity (km\,s$^{-1}$)}} 
	\end{picture}
	\label{fig_mgii_zp7}
\end{figure*}
\section{Quasar absorption line systems}\label{section_absline}
There are several intervening absorption line systems in the sightline of this quasar which is at emission redshift $z_{\rm em}=2.139$ \citep{Sargent88}. The combination of the Keck/HIRES and X-Shooter spectra provides us with powerful means to study these absorbers. The two absorbing systems at $z=0.38$ and $z=0.78$ were already known from previous studies of this sightline \citep{Sargent88_MgII,Guillemin97,Ellison05b,Rao06,Kacprzak07,Rahmani18}. While we find absorber systems over the redshift range from $z=0.38$ to $z=2.13$ we limit this study to those at $z<1.5$ as at larger redshifts the [\OII]\,$\lambda\lambda$\,3727,3729 (or simply [\OII] in the remaining of the paper) emission line falls out of the MUSE wavelength coverage.
 
To model the absorption lines we use the Voigt profile fitting code  \textsc{vpfit}\footnote{http://www.ast.cam.ac.uk/$\sim$rfc/vpfit.html} v10.0. In our Voigt profile model we assume a given component to have the same redshift and broadening parameter for all transitions. We relax this constraint when we model \CIV\ absorption profiles. The gas associated with \CIV\ absorption lines may not be co-spatial with that of singly ionized species. {The X-Shooter and Keck/HIRES data are in vacuum and hence to obtain the redshifts and velocities we use the vacuum rest frame wavelengths of different transitions.}

In the following we present the absorption line analysis of the absorbing system at $z=0.78$. Analysis of the remaining systems in this sightline is summarized in appendix \ref{appendix_absline}. We also exclude the system at $z=0.38$ from the following as the detailed study of this system was reported in our recent work \citep{Rahmani18}.
\begin{table*}
	\caption{Metal absorption line properties of the LLS at $z=0.78$ as extracted from the Voigt profile modeling of the Keck/HIRES spectrum of the quasar.}
	\begin{tabular}{lcccccc}
		\hline
		component& (1)&(2)&(3)&(4)\\
		\hline
		velocity (\kms)&$-78$&$-103$&$-260$&$-296$\\
		b (\kms) &$5.2\pm0.3$&$10.1\pm2.1$&$2.2\pm1.1$&$5.9\pm1.1$\\
		$\log$[N(\MgII)/cm$^{-2}$]&$\gtrsim14.4$&$12.28\pm0.07$&$12.17\pm0.10$&$12.22\pm0.05$\\
		$\log$[N(\FeII)/cm$^{-2}$] &$13.48\pm0.04$ &$<12.2$&$<11.9$&$<12.0$\\
		\hline
		\multicolumn{3}{l}{total column density of \MgII: log[N(\MgII)/cm$^{-2}$] $\gtrsim14.4$} \\
		\multicolumn{3}{l}{total column density of \FeII: log[N(\FeII)/cm$^{-2}$] $=13.48\pm0.04$} \\		
		\multicolumn{3}{l}{estimated dust-uncorrected metallicity [Fe/H]$=-1.1\pm0.3$}\\
		\hline
	\end{tabular}
	\label{tab_absorption_z0p78}
\end{table*}
\subsection{The LLS System at \zabs=0.780}\label{section_absline_zp78}
\citet{Rao06} used the HST/FOS spectrum of this quasar to deduce $\log$[\nhi\,/cm$^{-2}$]$=18.9^{+0.14}_{-0.11}$ for the LLS at $z=0.78$. Higher quality reduced spectra of HST/FOS observations have been produced by the more recent data reduction pipelines and can be obtained from the MAST\footnote{https://mast.stsci.edu} database (private communication, S. Rao). Hence, we downloaded the HST/FOS spectrum of this quasar and performed an analysis to measure the $N$(\HI) associated with this LLS. The spectrum covers Lyman series absorption lines and also the Lyman limit though with a poor spectral resolution of $\sim7$\,\AA. Our analysis resulted in a $\log$[\nhi\,/cm$^{-2}$]$=19.1\pm0.3$. We note that our measurement remains consistent with the value reported by \citet{Rao06} albeit with a larger uncertainty. 

In Fig.  \ref{fig_mgii_zp7} we present the Keck/HIRES spectrum of this quasar close to the \MgII\,$\lambda\lambda$2796,2803 and \FeII\,$\lambda$2344,2374,2586,2600 absorption lines at $z=0.780$. The zero velocity is set to $z=0.78025$ that is the systemic redshift of the host galaxy absorber (see section \ref{section_host_gal_zp78}). The absorption profiles are relatively simple composed of a strong component at $v=-78$\,\kms\ with a weak satellite one at $v=-103$\,\kms\ that is only detected in \MgII. There are two other well-separated optically thin components at $v=-260$\,\kms\ and $v=-296$\,\kms\ that are detected in \MgII\ but not in \FeII. The best fitting Voigt profiles to the absorption lines are shown as red continuous lines in Fig. \ref{fig_mgii_zp7}. Since, the main component of the \MgII\ is saturated we can only put a lower limit on its column density. The model parameters are summarized in Table \ref{tab_absorption_z0p78}. We further estimate the $3\sigma$ upper limits for the column density of \FeII\ for undetected components.

Based on the measured column density of \FeII\ we obtain [Fe/H]$=-1.1\pm0.3$ ([X/H]=$\log$(X/H)$-\log$(X/H)$_\odot$) where for the solar abundance we have used the value of $\log$(Fe/H)$_\odot=-4.5$ \citep{Asplund09}. The large uncertainty is due to the large error in the measured \HI\ column density. For systems at $\log$[\nhi\,/cm$^{-2}$]\,$\sim19$ some fraction of \HIII\ and \FeN\ are in \HII\ and \FeIII. \citet{Dessauges-Zavadsky03} demonstrated that even in the presence of such ionized fractions the \FeII/\HI\ ratio traces the abundance of Fe. Hence, we assume the ionization effect will not impact much our metallicity estimate. While \FeN\ is observed to be depleted into dust in the ISM of our Galaxy and also DLAs \citep{Vladilo98,Ledoux02a,Khare04,Wolfe05,Jenkins09,Kulkarni15,Jenkins17} several studies have shown that gas associated with LLSs is dust-poor \citep{Menard12,Lehner13,Fumagalli16LLS}. On contrary, \citet{Meiring09} found Fe depletion of $\gtrsim1$\,dex for two absorbers at similar N(\HI). Therefore, while we can not rule out the presence of the dust in this LLS, without measurements of un-depleted elements like Zn we cannot quantify how dust can affect the estimated metallicity. 

{If dust is associated with the gas in this LLS it can redden the spectrum of the quasar. We measure the reddening of the quasar spectrum induced by the foreground absorber by fitting a spectral template to the observed spectrum. For this purpose we use the template by \citet{Selsing16} and apply reddening assuming the extinction curve derived for the Small Magellanic Cloud \citet{Gordon03}. In order to better constrain the fit, we include near-infrared photometry retrieved from archival data from 2MASS: $J=16.74\pm0.06$\,mag, $H=16.64\pm0.08$\,mag and {\it WISE}: $W_1 = 16.71\pm0.03$\,mag. We do not include the $K$ band as is it contaminated by strong H$\alpha$ emission from the quasar, which we cannot constrain. Similarly, we do not include the remaining {\it WISE} photometric data as they are not covered by the spectral template. By fitting the photometry and spectrum simultaneously, we obtain a best-fit reddening of $E(B-V) = 0.01\pm0.01$\,mag (for a slightly redder intrinsic spectral slope of the template used in the fit: $\Delta\beta=0.03\pm0.06$). For details about the fitting procedure and the definition of $\Delta\beta$, see \citet{Krogager_thesis}.}

{Since there are several absorption systems along the line of sight, the measured reddening is a strict upper limit associated with the absorption system at $z_{\rm abs}=0.78$. Hence, the reddening from the absorber is negligible. This can be a tentative indication of lack of dust and depletion in this LLS.}
%
\begin{figure}
	\centering	
		\includegraphics[width=.45\textwidth,bb=0 130 595 712,clip=,angle=0]{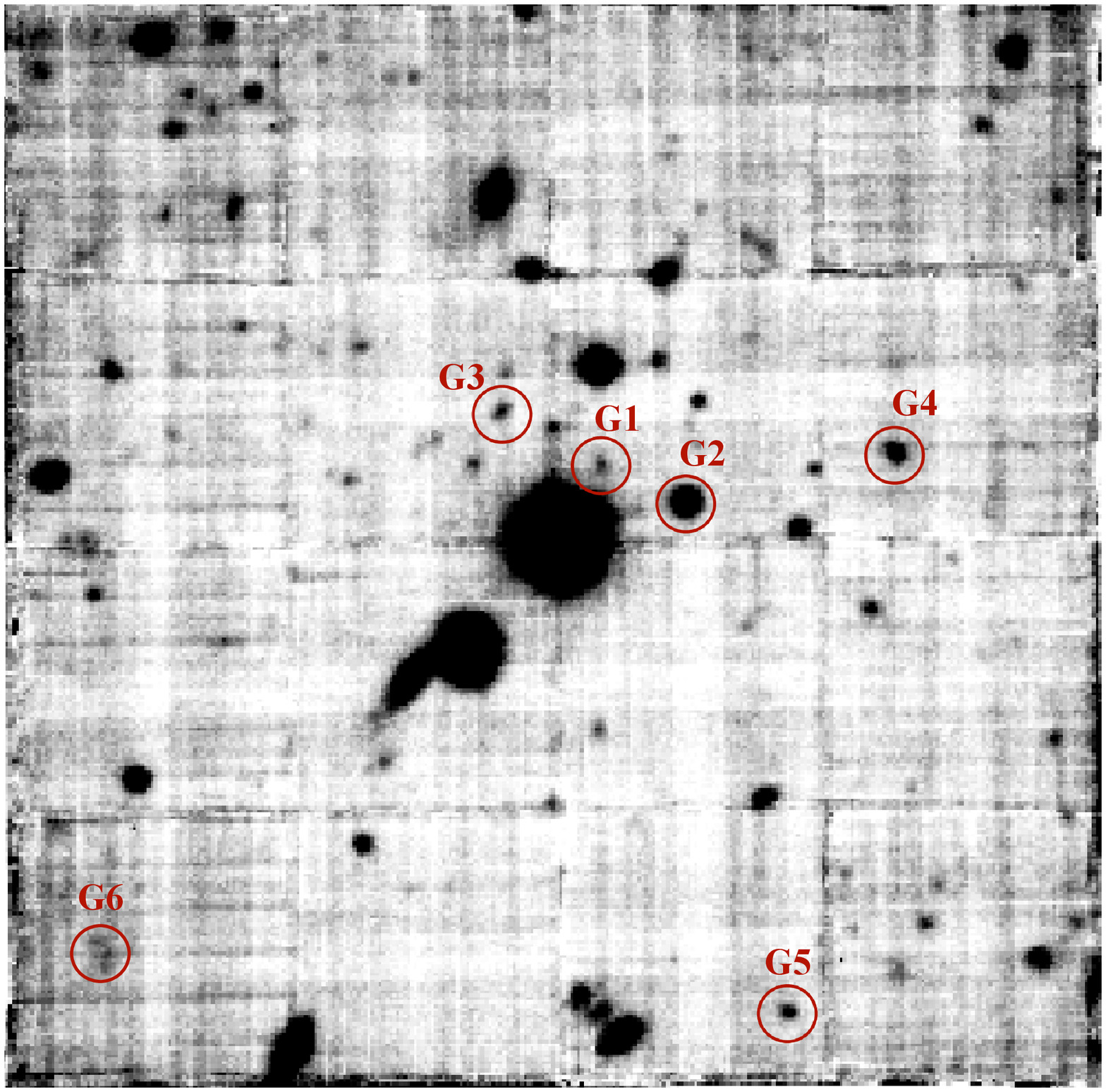}
	\caption{The $1'\times1'$ MUSE white light image of the field of Q0152$-020$. The ``G1'' to ``G6'' letters mark the sky position of the galaxies studied in this paper, ordered by impact parameter. ``G2'' is associated with the absorber at \zabs=0.780, ``G1'', ``G3'' and ``G4'' are associated with the absorber at $z=1.104$ and ``G5'' and ``G6'' are associated with the absorber at \zabs=1.211. The quasar resides in the centre of the field as indicated by \textbf{Q}.}
	\begin{picture}(0,0)(0,0)
	\put(-5,200){ \textcolor{red}{ \bf\large  Q}} 
	\end{picture}	
	\label{fig_wli}
\end{figure}
\section[]{Host galaxy absorbers} \label{section_host_gal}
We searched for the [\OII] nebular line emission at the redshift of each absorbing system in our MUSE datacube. Knowing the absorption redshifts we created narrow band images at the expected observed wavelength of [\OII] emission of each system and searched for the emitters. Our MUSE observation is deep enough to detect emitters with total emission line fluxes of $\gtrsim0.5\times10^{-17}$\,ergs\,cm$^{-2}$\,s$^{-1}$. To be complete we also inspected the spectra of all galaxies detected only in the continuum. We note that galaxies with AB magnitudes fainter than $m_r\sim23$\,mag will not have enough SNR for redshift measurements based on their absorption features. It is worth mentioning that possible purely continuum emitting galaxies (without emission lines) below the PSF of the quasar remain undetected.

In the following we discuss the analysis of the host galaxy absorber at $z=0.78$. Galaxies associated with the remaining absorbing systems are presented in appendix \ref{appendix_hostgal}. In Fig. \ref{fig_wli} we have marked the sky position of all the galaxies studied in this work over the MUSE white light image. 
\begin{figure*}
	\centering	
	\includegraphics[width=1.\textwidth,bb=-594 216 1206 576,clip=,angle=0]{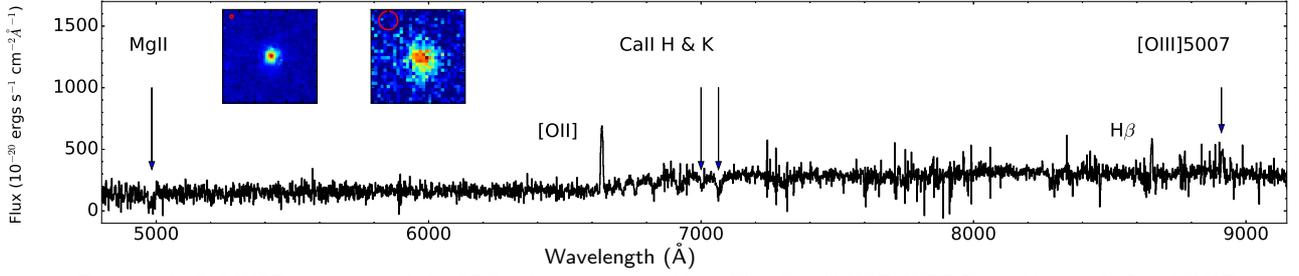} 
	\caption{Extracted 1D MUSE spectrum of the LLS-galaxy at $z=0.78025$. The $6"\times6"$ HST/WFPC2 $F702W$ and [\OII] MUSE narrow band postage stamp images of this galaxy are shown on the top left. {A circle showing the FWHM of the seeing is included in each panel.} We also detect the \MgII\,2796,2803 absorption lines from the host galaxy.}
	\begin{picture}(0,0)(0,0)
\put(-40,47){ \rotatebox{0}{  \sffamily Wavelength (\AA)}} 
\end{picture}	
	\label{fig_spec_full}
\end{figure*}
\subsection{Host galaxy absorber at \zabs=0.78}\label{section_host_gal_zp78}
We find one galaxy, G2, close to the absorption redshift at \zabs=0.78. Fig. \ref{fig_spec_full} presents the 1D spectrum of this galaxy extracted from the MUSE data cube. Strong nebular emission lines of [\OII], H$\beta$ and [\OIII]5007 are significantly detected. [\OIII]4958 is not detected but is consistent with the noise level of the flux. We use the standard deviation of the flux at the expected position of this line to find its flux upper limit. Table \ref{tab_flux_sfr_met} presents the measured fluxes of different emission lines from this galaxy, including dust corrections, calculated as explained in section \ref{EB-V} below. {We also observe Balmer absorption at the wavelength of H-$\beta$. Hence, to correctly estimate the emission line flux, we simultaneously model the profile with two Gaussian profiles to account for absorption and emission lines.} 

Interestingly we also detect in absorption the \MgII$\lambda\lambda$\,2796,2803 doublet that can be a tracer of outflowing or inflowing gas depending on their velocity with respect to the host galaxy \citep{Rubin12,Martin12}. However, the SNR in the continuum is not sufficient to study the \MgII\ absorption towards the host galaxy absorber.  
\begin{table}
	\small
	\caption{Extracted properties of G2, the host galaxy absorber at $z=0.78$.}
	\begin{tabular}{lccccc}
		\hline		\hline
		RA (J2000)&      28.111308\\
		Dec (J2000)&$-$20.017883\\
		impact parameter (kpc)& 54\\
		$z^1$&$0.78025\pm0.00007$\\
		\\
		$E(B-V)$ (mag)&$0.6\pm0.1^{2}$\\
		\\
		dust-corrected fluxes ($10^{-17}$\,ergs\,s$^{-1}$\,cm$^{-2}$)\\
		f([\OII]) &$68\pm19$\\
		f(H$\beta$)& $19\pm4$\\
		f([\OIII]$\lambda4958$) & $<3.5$ \\
		f([\OIII]$\lambda5007$) & $11\pm3$ \\
		\\
		dust-corrected SFR and emission metallicity  \\
		\\
		SFR$_{\rm H\beta}$ ($M_\odot$\,yr$^{-1}$) & $7\pm2$\\
		SFR$_{\rm \OII}$ ($M_\odot$\,yr$^{-1}$) & $13\pm5$\\		
		Z$_{\rm em}$ (Z$_\odot$) & $0.9\pm0.3$\\
		metallicity gradient (dex\,kpc$^{-1}$)& $-0.03$\\
		\hline
	\end{tabular}
\begin{flushleft}
	$^1$ Obtained by averaging the redshifts from H$\beta$ and [\OIII]5007 emission lines.\\
	$^2$ See Section \ref{EB-V} for the details of the derivation of $E(B-V)$.
\end{flushleft}
\label{tab_flux_sfr_met}
\end{table}
\subsubsection{Morphology from HST/WFPC2 broad-band imaging}
We modeled the HST/WFPC2 $F702W$ image of the host galaxy absorber at $z=0.78$ using the surface brightness fitting tool \textsc{galfit} \citep{Peng10AJ}. Using a single S\'{e}rsic model we obtained a good fit with a residual image ($\equiv$ [model]$-$[data]) to be consistent with the background noise level with no significant feature. Hence, this galaxy does not host any bar or bulge like structure and has not experienced a recent strong interaction with possible galaxy neighbors.

The model parameters are summarized in Table \ref{tab_HST_morphology}. The S\'{e}rsic index of such a model, $n=1.45$, indicates this galaxy has a disk dominated morphology. We note this galaxy has a disk with a half light radius of $0.34"\pm0.01"$ ($\equiv2.6\pm0.1$ kpc at $z=0.78$) and an inclination of $i=47\pm1\degree$.
\begin{table}
	\small
	\caption{Derived morphological parameters from the HST/WFPC2 $F702$ image for the LLS-galaxy.}
	\begin{tabular}{lccccccccccccc}
		\hline
		$r_{1/2}$ &  $n$ & $\sin i$ & PA \\
		(kpc)&&&(degree)\\
		\hline
		$2.6\pm0.1$  &  $1.45\pm0.07$ & $0.74\pm0.01$ & $-2\pm2$  \\
		\hline
	\end{tabular}
	\label{tab_HST_morphology}
	\begin{flushleft}
		Note. The columns of the table are: (1) half light radius; (2) S\'{e}rsic index; (3) inclination; (4) position angle \\
	\end{flushleft}
\end{table}
\begin{figure*}
	\centering
	\centerline{\includegraphics[width=1.\textwidth,bb=-238 76 850 715,clip=,angle=0]{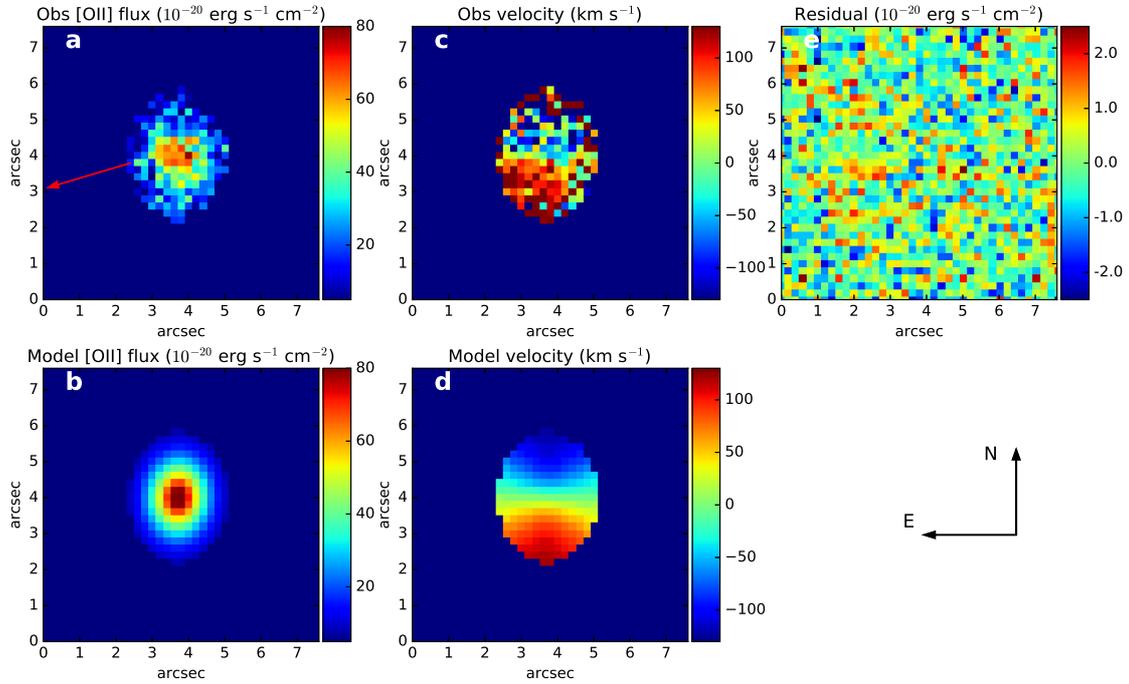}}
	\caption{Observed and modeled morphokinematics of G2. (a) Observed flux map of [\OII]. The red arrow indicates the direction towards the quasar. Both axes are in arcsec units where at $z=0.78$ the scale is 7.5 kpc/arcsec. The map presents a smooth disk-like distribution of [\OII] emission without noticeable distortions. (b) The modeled flux of [\OII] obtained using \textsc{galpak$^{3D}$}. (c,d): The observed and modeled velocity maps corresponding to the achieved model using \textsc{galpak$^{3D}$}. (e) The residual map (data-model) that shows the quality of the best model. The best obtained morphokinematic model has a half light radius of $r_{1/2}=6.7\pm0.2$\,kpc and a rotation curve with the maximum velocity of $V_{\rm max}=(2.0\pm0.1)\times10^2$\,\kms. These translate into $M_{\rm dyn}=(6.4\pm0.9)\times 10^{10}$\,M$_\odot$ and $M_{h}=(1.8\pm0.4)\times 10^{12}$\,M$_\odot$ (See section \ref{Morphokinematic} and eq. \ref{m_dyn} and \ref{m_halo} for the derivation of these quantities). The parameters extracted from the model are summarized in Table \ref{tab_galpak$^{3D}$}.}
	\label{fig_morph_f1}
\end{figure*}
\begin{table*}
	\centering
	\caption{Morphokinematic parameters of G2 obtained from 3D analysis of the [\OII] emission line using \textsc{galpak$^{3D}$}.}
	\begin{tabular}{lcccccccccc}
		\hline
		ID&$r_{1/2}$ & $\sin i^{1}$ & position angle & azimuthal angle      & $V_{\rm max}$ & $M_{\rm dyn}$ & $M_{\rm h}$         & $R_{\rm vir}$      & $M_{\rm \star}^{2}$\\
		&(kpc)        &             & (degree)          &(degree)                   & (\kms)            &  ($10^{10}$ \msun)& ($10^{12}$ \msun)& (kpc)&  ($10^{10}$\msun)\\
		(1)&(2)&(3)&(4)&(5)&(6)&(7)&(8)&(9)&(10)\\
		\hline
		G2&$6.7\pm0.2$ & 0.74  & $-2\pm3$  & $108\pm3$ & $(2.0\pm0.1)\times10^2$ & $6.4\pm0.9$ & $1.8\pm0.4$ & $(1.9\pm0.1)\times10^2$ & $3.6\pm1.5$\\  		
		\hline
	\end{tabular}
	\begin{flushleft}
		Note. The columns of the table are: (1) galaxy ID; (2) half light radius; (3) $\sin$ of inclination angle; (4) position angle; (5) azimuthal angle; (6) maximum rotation velocity; (7) dynamical mass; (8) halo mass; (9) virial radius; (10) total stellar mass. \\
		{$^1$} The inclination was fixed at the value obtained from the HST/WFPC2 image of the galaxy.\\
		{$^2$} $M_\star$ has been estimated using the Tully-Fisher relation \citep{Puech08}.
	\end{flushleft}
	\label{tab_galpak$^{3D}$}
\end{table*}
%
%
\subsubsection{Galaxy Kinematics}\label{Morphokinematic}
To study the morphokinematics of this galaxy we modeled the [\OII] emission line using \textsc{galpak$^{3D}$} \citep{Bouche15}. \textsc{galpak$^{3D}$} is a \textsc{python}-based optimization tool that simultaneously models the flux and kinematics of the emission line using a Monte Carlo Markov Chain (MCMC) algorithm. {To obtain a set of robust galaxy parameters, deconvolved from seeing, \textsc{galpak$^{3D}$} requires the shape of the PSF which is best modeled with Moffat function as discussed in section \ref{section_observation}.} While modeling the morphokinematics of the [\OII] emission line we assumed the wavelength ratio of the doublet to be fixed at its theoretical value of 0.9993 \citep{Wiese96}. \citet{Contini16} demonstrated that at small inclination ($i\lessapprox50\degree$) \textsc{galpak$^{3D}$} may underestimate the inclination of galaxies. Hence, to obtain a more robust parameter measurements from \textsc{galpak$^{3D}$}, we fixed the inclination at $i=47\degree$ as measured from the high spatial resolution HST/WFPC2 image. 
{We allow a 10000-long MCMC to obtain a robust sample of the posterior probability. Also to check the convergence of the MCMC chain we inspect the possible correlation between each pair of parameters in their 2D distributions}. 
%

 Fig. \ref{fig_morph_f1} shows the best morphokinematic model for G2 that is achieved using an exponential light profile along with an \textit{arctan} rotation curve. This model corresponds to a half light radius of $r_{1/2}=6.7\pm0.2$\,kpc and a maximum rotation velocity of $V_{\rm max}=(2.0\pm0.1)\times10^2$\,\kms. We estimate the enclosed mass within half light radius using the following equation:
 \begin{equation}
 M_{\rm dyn}(r<r_{1/2}) =  r_{1/2}V_{\rm max}^{2}/G
 \label{m_dyn}
 \end{equation}
 to be $(6.4\pm0.9)\times10^{10}$\,\msun. Assuming a spherical virialized collapse model \citep{Mo98}:
 \begin{equation}
 M_{h}=0.1 H(z)^{-1}G^{-1} V_{\rm max}^3
 \label{m_halo}
 \end{equation}
 where $H(z)=H_0 \sqrt[]{\Omega_\Lambda + \Omega_m(1+z)^3}$, we find the halo mass of this galaxy to be $M_h=(1.8\pm0.4)\times10^{12}$\,\msun. From such a collapse model we also obtain the $R_{\rm vir}=0.1V_{\rm max}/H(z)=(1.9\pm0.1)\times10^2$\,kpc. Therefore, the sightline to the quasar, where the absorption is produced, is located at $\sim$0.3\,$R_{\rm vir}$.
 
 The measured $r_{1/2}$ of [\OII] emission line obtained using \textsc{galpak$^{3D}$} is $2.5\pm0.1$ times larger than $r_{1/2}$ obtained for the continuum using \textsc{galfit} (see Table \ref{tab_HST_morphology}).{ To check if this difference is not due to the different modeling procedures or datasets, we extracted continuum narrow band image of the galaxy from our MUSE data. Next we used \textsc{galfit} to model this image where we found $r_{1/2}=3.0\pm0.2$\,kpc. This is a bit larger than the $r_{1/2}$ obtained from HST image but still consistent within 2$\sigma$. Hence, the difference between the size of the continuum emitting region and emission line region is not due to systematic errors in our measurements. We further confirm this conclusion by modeling the [\OII] narrow band image (extracted from MUSE) using \textsc{galfit} to find $r_{1/2}=6.2\pm0.4$\,kpc (which is consistent with the $r_{1/2}$ from \textsc{galpak$^{3D}$}). 
 Rather, it appears that the gaseous disk of this galaxy is significantly larger than the stellar disk \citep[e.g., see][for similar results but for larger samples of galaxies]{Koopmann06,Bamford07,Puech10}. While recent merger and strong SNe can be considered as two possible scenario to provide the [\OII] emitting gas surrounding galaxies the real picture demands a study using a much larger smaple of galaxies [Contini et al. (in prep.)].}  
 
There exists a tight correlation between the stellar mass and the maximum rotation velocity of galaxies, known as the stellar-mass Tully-Fisher relation \citep{Tully77}. We use the reported Tully-Fisher relation for a sample of galaxies at $z\sim0.6$ \citep{Puech08} and $V_{\rm max}=(2.0\pm0.1)\times10^2$\,\kms to deduce the stellar mass of this galaxy $M_\star=(3.6\pm1.5)\times10^{10}$\,\msun\ (or $\log$($M_\star/M_\odot$)$=10.6\pm0.2$). { Therefore, the stellar-mass to dark matter fraction for this halo is $2.0\pm0.9\times10^{-2}$. This is in agreement with the stellar to dark matter fraction obtained using the technique of ``abundance matching'' which is $\sim0.02$ \citep[e.g.,][]{Behroozi10}}.
%
%
%
\subsubsection[EB-V]{Extinction correction}\label{EB-V}
%
The Balmer decrement (f(H$\alpha$)/f(H$\beta$)) is usually used to measure the intrinsic dust extinction, E(B-V), in star forming galaxies. Since H$\alpha$ from this galaxy is not covered in the MUSE data cube we can not directly measure the E(B-V). Alternatively, \citet{Garn10} used a large sample of galaxies in the local Universe to calibrate the total extinction for H$\alpha$, $A_{H\alpha}$, versus the stellar mass of galaxies:
\begin{equation}
A_{H\alpha}=\sum_{i=0}^{3} B_i X^i
\end{equation}
where $X=\log_{10}{\rm(M_\star/10^{10}M_\odot)}$ and $B_i$ are the polynomial coefficients. \citet{Sobral12} later demonstrated that the same calibration is valid at much higher redshifts, $z\sim1.5$. Assuming that absorption selected samples have identical dust properties to flux limited samples, 
we use the same calibration and the stellar mass of this galaxy, $\log_{10}(M_\star/M_\odot)=10.6\pm0.2$,  to estimate the $A_{H\alpha}=1.36\pm0.14$. Adopting a Small Magellanic Cloud-type (SMC-type) extinction law we measure the $E(B-V)=0.6\pm0.1$. With this value of $E(B-V)$ and the SMC extinction curve we calculate extinction corrections to the line fluxes of [\OII], H$\beta$ and [\OIII]. The values listed in Table \ref{tab_flux_sfr_met} include such corrections. 

To obtain the dust-corrected star formation rate (SFR) we first convert the H$\beta$ flux into H$\alpha$ flux according to the theoretical Balmer decrement (2.88) in a typical \HII\ region. Therefore we estimate an H$\alpha$ flux of $F{(\rm H\alpha)}=(54\pm12)\times10^{-17}$\,ergs\,s$^{-1}$\,cm$^{-2}$. We convert the total H$\alpha$ flux of this galaxy to SFR assuming a \citet{Kennicutt98} conversion corrected to a \citet{Chabrier03} initial mass function. Using such a conversion we find the dust-corrected SFR of this galaxy to be $7\pm2$\,$M_\odot~{\rm year^{-1}}$. We also estimate the dust-corrected SFR based on the [\OII] emission using the calibration given by \citet{Kewley04} to be $13\pm5$\,M$_\odot$\,yr$^{-1}$ which is larger than the SFR obtained based on H$\beta$ but still marginally consistent. For the remaining of the paper we consider the SFR of these galaxy to be the average of these two estimates, $10^{+8}_{-5}$\,M$_\odot$\,yr$^{-1}$. The quoted error is enough conservative to also include uncertainties via utilizing different extinction laws. 
\subsubsection{Emission metallicity}
We deduce the oxygen abundance, O/H, of the ionized gas from the dust-corrected flux ratios of different nebular emission lines based on $R23$ index \citep{Kobulnicky99}. We find the lower and upper branch values of O/H based on $R23$ as calibrated by \citet{Kobulnicky99} to be $7.9\pm0.2$ and $8.6\pm0.2$. Such abundances translate to [O/H]=$-0.8\pm0.2$ and [O/H]=$-0.1\pm0.2$ [12+log(O/H)$_\odot$=8.69, \citealt{Asplund09}]. However, there exists a tight correlation between the stellar mass and the metallicity of galaxies \citep[e.g.,][]{Tremonti04,Savaglio05,Queyrel09,Zahid11,Foster12,Yabe15,Bian17}. Several studies have shown that galaxies with stellar masses of $\gtrsim10^{10.5}$\,$M_\odot$ do not have metallicities less than [O/H]$\sim-0.2$ \citep[e.g.,][]{Tremonti04,Zahid11}. Therefore, based on the stellar mass of this galaxy ($M_\star=10^{10.6}$\,$M_\odot$) the [O/H] inferred from the upper branch, [O/H]$=-0.1\pm0.2$ is the preferred value for this galaxy. 

Using $R23$ (and assuming we are on the upper branch) we can measure O/H at different galactocentric distances in the disk and we deduce a gradient of $\sim-0.03$\,dex\,kpc$^{-1}$. This value is consistent with the typical $\sim-0.02$\,dex\,kpc$^{-1}$ measured for the host galaxies of strong \HI\ absorbers \citep[e.g.,][]{Moller13,Christensen14,Peroux14,Rahmani16,Rahmani18}. Either value is consistent with $\sim1$\,dex difference between emission-based O/H and absorption-based Fe/H at an impact parameter 54\,kpc, if the gradient continues at this rate all the way to 54 kpc. While this agreement is suggestive, we do note that later (section \ref{disk_scenario}) we discount an extended disk origin for the absorber on kinematic grounds. 


In comparing the absorption and emission abundances we have assumed that O/Fe is solar. This may be an invalid assumption in this starforming galaxy, since the generation of core-collapse SN from massive stars can lead to O/Fe values that are 2--3 times larger than that of solar. We note however, this is still not enough to explain the observed difference between the abundances of the two gas phases. However, it is worth noting that the effect of dust depletion can be large to explain the observed metallicity difference. However, as noted earlier we can not quantify it in this absorber.
%
%
%
\section{Nature of the absorbing gas at \zabs=0.78}\label{section_conclude}
We have detected the host galaxy absorber of an LLS at $z=0.78$ at an impact parameter of $b=54$\,kpc. In the following we study the possible nature of this LLS.
\subsection{Galaxy disk}\label{disk_scenario}
This LLS may originate from the gas associated with the extended disk of the galaxy. This scenario is supported by the disk-like morphology of the galaxy and the fact that the extrapolated metallicity of the galaxy is consistent with the absorption metallicity. In such a scenario one expects the LLS line of sight velocity to match the one predicted from the rotation curve of the galaxy (\citealt{Chen05}; but see \citealt{Neeleman16_CO} for a counter-example). As indicated in panel \textbf{c} of Fig. \ref{fig_morph_f1} the rotation of the disk towards the quasar sightline would produce a redshifted velocity with respect to the systemic velocity of the galaxy. To be more precise, we use the best morphokinematic model of the galaxy obtained from \textsc{galpak$^{3D}$} and extrapolate it to the position of the quasar. The projected velocity at the apparent position of the quasar is $v_{Q}=50\pm10$\,\kms. Given the large velocity differences ($\delta v>130$\,\kms) between the disk and absorption components (located at $v \sim -80$\,\kms\ and $v \sim -300$\,\kms--see section \ref{section_absline_zp78}), we conclude that this LLS is not associated with the disk of the galaxy.

Azimuthal angle, $\phi$, in quasar-galaxy pairs is defined to be the angle between the galaxy major axis and the quasar line of sight. At $\phi\approx90\degree$ the sightline towards the quasar probes the galaxy poles while at $\phi\approx0\degree$ it passes close to the galactic disk. We have measured $\phi=70\degree$ for this quasar-galaxy pair that shows the sightlines probes the regions far from the galaxy disk. This consideration further argues against the disk scenario as the possible origin of the absorbing gas.
\subsection{Warped disk}\label{accretion_scenario}
Warped extended disks, that can be inclined up to $\sim20\degree$ with respect to the stellar disk, are commonly observed to be associated with disk galaxies in the local Universe \citep[e.g.,][]{Briggs90,Garcia-Ruiz02,Heald11}. Galactic warps are usually detected at a few effective radii but in extreme cases have also been seen to extend up to $\sim60$\,kpc. There also exists observational evidence of warped disks at higher redshifts \citep{Bouche13,Burchett13,Diamond-Stanic16,Rahmani18}. Gas accretion is considered as one of the most plausible explanation for the presence of such structures \citep{Stewart11b,Shen13,Stewart17}. Due to the large velocity difference of the LLS with respect to the disk and the azimuthal angle of $\phi=70\degree$ we reject such a scenario for the nature of the LLS as galactic warps are expected to corotate with the galaxy disks, though with only a small lag. 
%
\subsection{High velocity clouds}\label{HVC_scenario}
High velocity clouds (HVCs) associated with the CGM of our Galaxy are defined as gaseous structures with $|v_{LSR}|\gtrsim90$\,\kms \citep{Wakker97ARA&A}. HVCs are also observed in the CGM of other galaxies in the Local group with similar \nhi\ and metallicities as this LLS \citep[e.g.,][]{Lehner10,Lehner12,Richter16}. The majority of the HVCs in the Milky Way reside within $b=15$\,kpc \citep{Wakker07,Wakker08} which is $\simeq4$ times smaller than the impact parameter of this LLS ($b=54$\,kpc). HVCs associated with tidal streams of the LMC have been found at larger distances of up to $\sim50$ kpc and with velocities as large as $300$\,\kms. In the case of M31, HVCs are observed up to distances of 150 kpc that may not be related to any tidal interactions \citep{Westmeier05,Westmeier07,Wolfe_s16}. Therefore, both the strong component at $v\sim-80$\,\kms and the two weaker ones at $v\sim-300$\,\kms\ may originate from HVC like structures in the CGM of this galaxy. 
%
\begin{figure}
	\centering	
		\vspace*{-1cm}
	\centerline{\vbox{
			\hspace*{-2cm}\includegraphics[width=1.0\hsize,bb=0 0 595 842,clip=,angle=270]{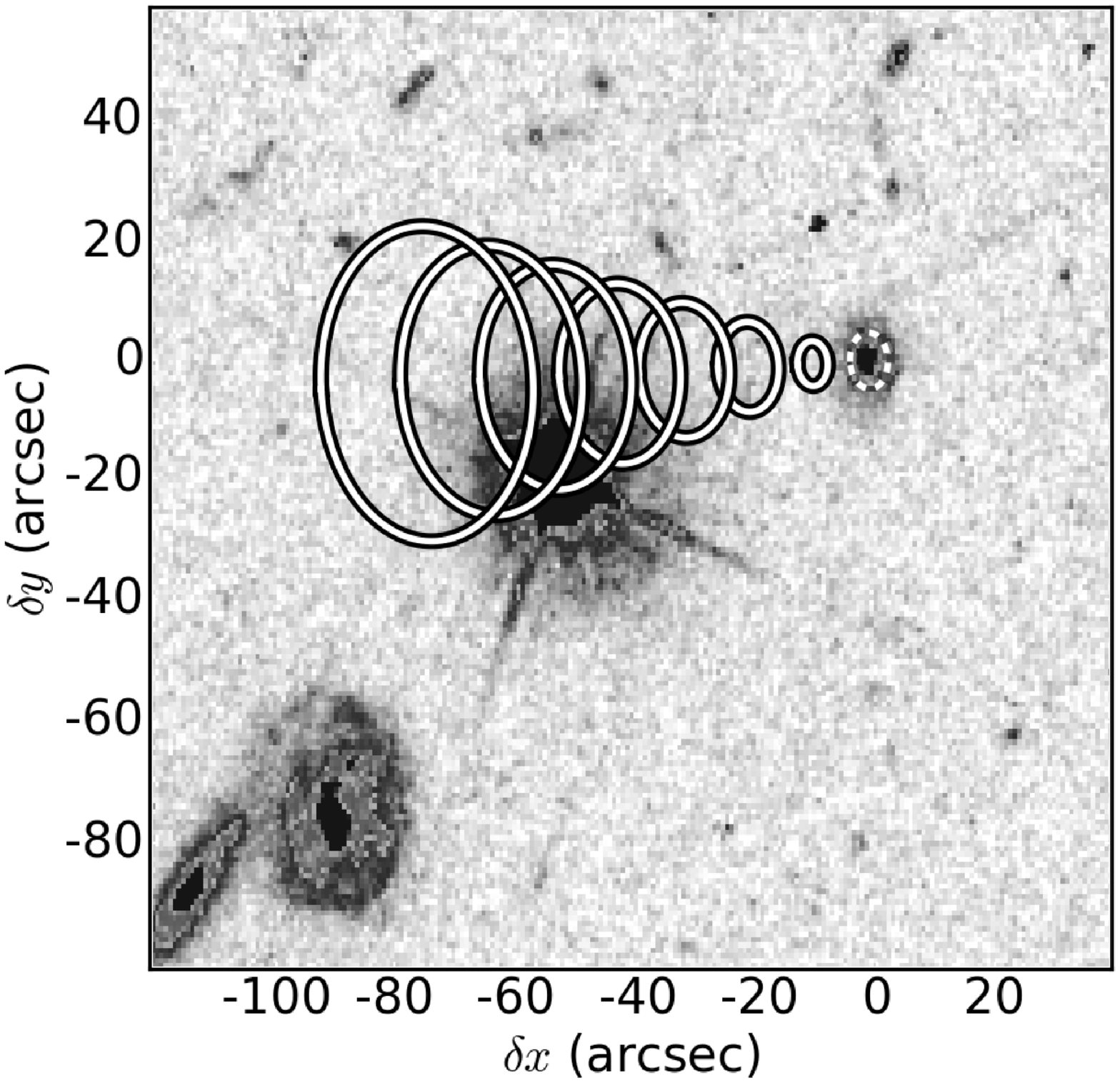}
			\includegraphics[width=0.9\hsize,bb=93 318 518 473,clip=,angle=0]{feii_2344_obs.ps}
			\includegraphics[width=0.9\hsize,bb=93 318 518 473,clip=,angle=0]{feii_2344_sim.ps}
	}}
	\caption{Representation of the cone model utilized to reproduce the quasar absorption line spectrum. \textit{top:} Zoomed HST/WFPC2 image over which the cone model is illustrated in the sky plane (xy). The dashed and inclined circles respectively indicate the galaxy disk and the wind cone. \textit{bottom} (B1): Normalized Keck/HIRES quasar spectrum close to the \FeII\,2344 absorption line. \textit{bottom} (B2): The reconstructed profile based on the galactic wind cone model. The best match to the observed spectrum is obtained with a wind velocity of $V_{\rm w}$=110\,\kms\ and a cone opening angle of $\theta_{\rm max}=15^\circ$.}
	\label{fig_wind_model}
\end{figure}
\subsection{Galactic wind}\label{wind_scenario}
 The azimuthal angle of $\phi=70^\circ$ between Q0152$-$020 and G2 indicates that the quasar sightline probes closer to the galactic pole where galactic winds are expected to reside \citep[e.g.,][]{Kacprzak12,Bouche12a}. By dividing the dust-corrected SFR of $10^{+8}_{-5}$\,M$_\odot$\,yr$^{-1}$ by the surface of $\pi r_{1/2}^{2}$, where $r_{1/2}=6.7\pm0.2$\,kpc, we measure the average SFR surface density of $\Sigma_{\rm SFR}=0.07^{+0.06}_{-0.04}$\,M$_{\odot}$\,yr$^{-1}$\,kpc$^{-2}$. This $\Sigma_{\rm SFR}$ is consistent with the threshold of $\Sigma_{\rm SFR}\gtrsim0.1$\,M$_{\odot}$\,yr$^{-1}$\,kpc$^{-2}$ required for driving large-scale galactic winds \citep{Heckman02,Newman12,Bordoloi14}. 
 
 In order to see if this LLS can be associated with a galactic wind we try to reproduce the absorption lines by utilizing a wind cone model as described in \citet{Bouche12a} \citep[and also in][]{Schroetter15,Schroetter16}. In such a wind model one assumes a cone, with an opening angle $\theta_{\rm max}$\footnote{$\theta_{\rm max}$ is defined from the central axis, and the cone subtends an area $\Sigma$ of $\pi \cdot \theta_{\rm max}^2$.}, filled with randomly distributed gas clouds decreasing in number as $1/r^2$, $r$ being the distance from the centre of the galaxy. The clouds in this model are quickly accelerated to their terminal velocity $V_{\rm w}$ in a few kpc ($<10$\,kpc).  The projection of the velocity of particles in the cone, onto the quasar sightline, provides us with an optical depth $\tau_v$ which is then turned into a simulated absorption profile. The comparison with the observed absorption line is finally carried out after convolving the simulated profile with the instrumental line spread function and adding to it Poisson noise  in order to reproduce the instrumental noise. 

We achieved the best match between the simulated and observed absorption profiles with a wind velocity of $V_{\rm w}=110\pm4$\,\kms\ and a cone opening angle of $\theta_{\rm max}=15\pm2^\circ$. We show a schematic representation of such a model in the \textit{top} panel of Fig. \ref{fig_wind_model} where the cone is oriented to be perpendicular to the disk of the galaxy. In the \textit{bottom} panels of Fig. \ref{fig_wind_model} we present the observed (panel B1) and simulated (panel B2) spectra of \FeII\,2344 which verifies that such a wind model can reproduce the main absorption component of the LLS. 

Using the wind velocity and the cone opening angle, we can now derive the ejected mass rate of the wind, $\dot M_{\rm{out}}$, by using Equation 1 of \citet{Schroetter15}:
\begin{equation}
\dot M_{\rm{out}} = \mu \cdot N_{\rm H} (b) \cdot b \cdot V_{\rm{w}} \cdot \frac{\pi}{2} \cdot \theta_{\rm{max}}
\label{eq_Mout}
\end{equation}
where $\mu$ is the mean atomic weight, $b$ the impact parameter, $\theta_{\rm max}$ the cone opening angle, $V_{\rm{w}}$ the wind velocity and $N_{\rm H}(b)$ the hydrogen column density at the distance $b$. Using Eq. \ref{eq_Mout} and assuming $\mu=1.5$ we find an $\dot M_{\rm{out}}=0.38^{+0.15}_{-0.10}$\,M$_{\odot}$\,yr$^{-1}$. Since some fraction of hydrogen in LLSs is ionized \citep[e.g.,][]{Dessauges-Zavadsky03,Lehner13,Fumagalli16LLS,Prochaska17} this mass ejection rate should be considered as a lower limit. Assuming a symmetry for the wind geometry (i.e. two cones of ejecting gas from both sides of the galaxy) the mass ejection rate would be twice that. From the value of the ejected mass rate of both cones, we can derive the loading factor $\eta$, defined as $\dot M_{\rm{out}}$/SFR. With the $\dot M_{\rm{out}}$ of $\gtrsim0.8$\,M$_{\odot}$\,yr$^{-1}$ we find a loading factor $\eta\gtrsim0.1$. 

We further derive the escape velocity $V_{\rm {esc}}(b)$ of the halo at a distance $b$ from the host galaxy assuming an isothermal sphere profile given by \citet{Veilleux05ARA&A}:
\begin{equation}
V_{\rm esc}=V_{\rm max}\times\sqrt{2\left[1+\ln \left(\frac{R_{\rm vir}}{r}\right)\right]}
\label{eq_Vesc}
\end{equation}
where $V_{\rm max}$ is the maximum rotation velocity of the galaxy and $R_{\rm vir}$ is the virial radius. With a $V_{\rm max}=(2.0\pm0.1)\times10^2$\,\kms\ and $R_{\rm vir}=(1.9\pm0.1)\times10^2$\,kpc, we find an escape velocity of $V_{\rm esc}\simeq430$\,\kms. This value is approximately four times larger than the wind velocity. Therefore, such a wind can not escape the gravitational potential of the halo. {The estimated escape velocity assuming a Navarro-Frenk-White (so called NFW) profile \citep{Navarro96_NFW} for the halo is 250\,\kms\ which is $\sim1.7$ times smaller than the value deduced above for an isothermal sphere. However, the main conclusion remains unchanged since such an escape velocity is still $\sim2.3$ times larger than the wind speed. }

We further explored if the wind cone model can explain the origin of the other, weaker, absorption components (see Fig. \ref{fig_mgii_zp7}). However, within a reasonable range of parameters for $V_w$ and $\theta_{\rm max}$ we were not able to reproduce such absorption profiles. Hence, the remaining components may not originate in a wind associated with this galaxy. 

It is worth noting that the measured metallicity from the emission lines is $\sim1$\,dex larger than that obtained from the gas seen in absorption. This appears to be inconsistent with the wind picture in which one expects to observe higher or equal metallicity gas in absorption compared to that of the host galaxy. However, the overproduction of O compared to Fe due to core-collapse SN and the dust depletion of Fe can reduce the metallicity difference significantly. Moreover, cosmological hydrodynamical simulations predict a significant mixing of the wind materials at large radii ($\gtrsim0.25R_{\rm vir}$) with the pristine accreted gas from IGM that significantly dilutes the wind metallicity \citep[e.g.,][]{Muratov17,Mitchell17}.
\section[summary]{Summary and conclusion}\label{section_summary}
In this work we have studied an LLS with $\log$[N(\HI)/cm$^{-2}$]=$19.1\pm0.3$ at $z=0.78$ towards Q0152$-$020. The metal absorption profiles associated with this LLS consist of a strong main component (seen in \MgII\ and \FeII) but also another two well separated weaker components at velocity differences of $\sim200$\,\kms\ (that are only detected in \MgII). We have measured a dust-uncorrected absorption metallicity of [Fe/H]$=-1.1\pm0.3$\,dex.

At the absorption redshift of \zabs=0.78 our VLT/MUSE data is sensitive to a dust-uncorrected SFR=0.5\,M$_\odot$\,yr$^{-1}$. Inspecting our VLT/MUSE data we find one galaxy at a velocity separation of $v=78$\,\kms\ with respect to the main absorption component. This galaxy is located at an impact parameter of $b=54$\,kpc. We confirm the redshift of this galaxy by detecting [\OII], [\OIII]5007 and H$\beta$ emission lines. 

By modeling the HST/WFPC2 $F702$ broad-band image of this galaxy using a S\'ersic profile we obtained a fit with $n=1.45\pm0.07$. The morphokinematic modeling, using the MUSE data cube, shows this galaxy has a rotation curve with a $V_{\rm max}=(2.0\pm0.1)\times10^2$\,\kms. Such analyses further lead us to conclude that this disky galaxy has not suffered from recent strong interactions with other galaxies. 

We infer the stellar mass using the Tully-Fisher relation to be $M_\star=(3.6\pm1.5)\times10^{10}$\,M$_\odot$. We further measure a dust-corrected SFR for this galaxy to be $10^{+8}_{-5}$\,M$_\odot$\,yr$^{-1}$ and oxygen abundance of [O/H]=$-0.1\pm0.2$. We also estimate a metallicity gradient of $\sim0.03$\,dex\,kpc$^{-1}$.

At the position of the quasar, the predicted velocity of gas corotating with the disk of the galaxy is +50\,\kms. Since the velocities of different absorbing components are $v\sim-80$\,\kms\ and $v\sim-300$\,\kms\ we reject a scenario in which the LLS originates from the stellar disk or a warped disk associated with this galaxy. 

HVCs, associated with the galaxies in the Local group, have similar metallicities and velocities to this LLS and can be found at comparable impact parameters to their host galaxies. Hence, we can not refute the possibility that one or more absorption components can be associated with HVC like clouds in the CGM of the host galaxy. 

We also explore the possibility that the LLS arises from a galactic wind phenomenon. This model is supported by the azimuthal angle of $\phi=70^\circ$ which shows the quasar sightline passes close to the galactic pole. We try to reproduce the absorption profile using a wind model where gas clouds are confined within a cone and have reached their terminal velocity. A good match to the absorption profile of one of the absorption components is achieved with a wind velocity of $V_{\rm w}=110\pm4$\,\kms\ and cone opening angle of $\theta_{\rm max}=15\pm2^\circ$. We estimate a total mass ejection rate of $\dot{M}\gtrsim0.8$\,M$_{\odot}$\,yr$^{-1}$ and a loading factor of $\eta\gtrsim0.1$. {This parameter plays a crucial role in reproducing the observed population of galaxies in simulations \citep[e.g.][]{Dutton12,Dave13,Muratov15,Hayward17}. However, our estimated value of $\eta\gtrsim0.1$ is in agreement with those from FIRE simulations \citep[][]{Muratov15}.} We further find the escape velocity from the halo of this galaxy is almost four times larger than the wind speed. Therefore, this outflowing material will not be able to leave the halo of this galaxy. 

\citet{Guillemin97} had also searched for the host galaxy of this absorbing system where they associated it with a galaxy at $z=0.603$ ($\Delta v\gtrsim3\times10^{5}$\,\kms). Thanks to the large field of view and the high sensitivity of MUSE we can now reliably identify and study the host galaxy absorber. Further combining MUSE data with the high spectral resolution quasar absorption line data and HST image of the field we further studied the nature of the absorbing gas. Altogether our results show the power of IFU observations of quasar galaxy pairs in CGM studies.
\section*{Acknowledgements}
HR thanks Francois Hammer, Mathieu Puech and Hector Flores for useful suggestions. 
CP was supported by an ESO science visitor programme and the DFG cluster of excellence `Origin and Structure of the Universe'. This work has been carried out thanks to the support of the OCEVU Labex (ANR-11-LABX-0060) and the A*MIDEX project (ANR-11-IDEX-0001-02) funded by the ``Investissements d'Avenir'' French government program managed by the ANR. This research was supported by the DFG cluster of excellence `Origin and Structure of the Universe' (www.universe-cluster.de). RA thanks ESO and CNES for the support of her PhD. VPK gratefully acknowledges partial support from a NASA/STScI grant for program GO 13801, and additional support from NASA grants NNX14AG74G and  NNX17AJ26G. The research leading to these results has received funding from the French Agence Nationale de la Recherche under grant no ANR-17-CE31-0011-01.
%




\appendix

\def\aj{AJ}%
\def\actaa{Acta Astron.}%
\def\araa{ARA\&A}%
\def\apj{ApJ}%
\def\apjl{ApJ}%
\def\apjs{ApJS}%
\def\ao{Appl.~Opt.}%
\def\apss{Ap\&SS}%
\def\aap{A\&A}%
\def\aapr{A\&A~Rev.}%
\def\aaps{A\&AS}%
\def\azh{AZh}%
\def\baas{BAAS}%
\def\bac{Bull. astr. Inst. Czechosl.}%
\def\caa{Chinese Astron. Astrophys.}%
\def\cjaa{Chinese J. Astron. Astrophys.}%
\def\icarus{Icarus}%
\def\jcap{J. Cosmology Astropart. Phys.}%
\def\jrasc{JRASC}%
\def\mnras{MNRAS}%
\def\memras{MmRAS}%
\def\na{New A}%
\def\nar{New A Rev.}%
\def\pasa{PASA}%
\def\pra{Phys.~Rev.~A}%
\def\prb{Phys.~Rev.~B}%
\def\prc{Phys.~Rev.~C}%
\def\prd{Phys.~Rev.~D}%
\def\pre{Phys.~Rev.~E}%
\def\prl{Phys.~Rev.~Lett.}%
\def\pasp{PASP}%
\def\pasj{PASJ}%
\def\qjras{QJRAS}%
\def\rmxaa{Rev. Mexicana Astron. Astrofis.}%
\def\skytel{S\&T}%
\def\solphys{Sol.~Phys.}%
\def\sovast{Soviet~Ast.}%
\def\ssr{Space~Sci.~Rev.}%
\def\zap{ZAp}%
\def\nat{Nature}%
\def\iaucirc{IAU~Circ.}%
\def\aplett{Astrophys.~Lett.}%
\def\apspr{Astrophys.~Space~Phys.~Res.}%
\def\bain{Bull.~Astron.~Inst.~Netherlands}%
\def\fcp{Fund.~Cosmic~Phys.}%
\def\gca{Geochim.~Cosmochim.~Acta}%
\def\grl{Geophys.~Res.~Lett.}%
\def\jcp{J.~Chem.~Phys.}%
\def\jgr{J.~Geophys.~Res.}%
\def\jqsrt{J.~Quant.~Spec.~Radiat.~Transf.}%
\def\memsai{Mem.~Soc.~Astron.~Italiana}%
\def\nphysa{Nucl.~Phys.~A}%
\def\physrep{Phys.~Rep.}%
\def\physscr{Phys.~Scr}%
\def\planss{Planet.~Space~Sci.}%
\def\procspie{Proc.~SPIE}%
\let\astap=\aap
\let\apjlett=\apjl
\let\apjsupp=\apjs
\let\applopt=\ao
\bibliographystyle{mnras}
\bibliography{/Users/hrahmani/work/IGM/files-ref/bib.bib}
%
%
%
%
%
%
%
%
\section{New X-Shooter observations}\label{XSH_OBS}
Q0152$-$020 has been observed with VLT/X-Shooter \citep{Vernet11} under the ESO program 095.A-0338 (PI: P\'{e}roux). The X-Shooter spectrograph covers a wavelength range of 0.3 $\mu$m to 2.3 $\mu$m at medium resolution in a simultaneous use of three arms in UVB, VIS and NIR. For a robust sky subtraction the nodding mode was used following the ABBA scheme with the default nod length (2.5$''$). Slit widths of 1.0$''$,0.9$''$ and 0.9$''$ were used for respectively UVB, VIS and NIR arms. These choices of slit widths result in formal spectral resolving powers of 5100, 8800 and 5600 for the UVB, VIS and NIR respectively.

To reduce the raw data we used the X-Shooter pipeline version 2.2.0\footnote{http://www.eso.org/sci/software/pipelines/} \citep{Goldoni11}. We followed a standard procedure of X-Shooter data reduction to obtain the 1D spectrum of the quasar \citep{Rahmani16}. We first made an initial estimate for the wavelength solution and position of the centre and edges of each order. Then we obtained the accurate position of the centre of echelle orders and followed this step by generating the master flat frame out of five individual lamp exposures. Next, the 2D wavelength solution was found and modified by applying a flexure correction. Finally, we subtracted each pair of nods to obtain the flat-fielded, wavelength calibrated and sky subtracted 2D spectrum. As we are primarily interested in studying the (narrow) absorption lines, we did not attempt to flux-calibrate the spectrum.
%
\section{Properties of other absorbing systems towards Q0152$-$020} \label{appendix_absline}
In this section we present the absorption line properties of absorbing systems towards Q0152$-$020 at $1<z<1.5$. In Table \ref{tab_other_abs} we summarize the column density measurements for different ions. We do not detect the Ly$\alpha$ absorption line to be associated with any of these absorbers. Hence, we estimated upper limits on $N(\HI)$ from the HST/FOS data. 
\subsection{Absorbing system at \zabs=1.104}
\begin{figure}
	\centering	
	\includegraphics[width=1.\hsize,bb=134 90 447 678,clip=,angle=0]{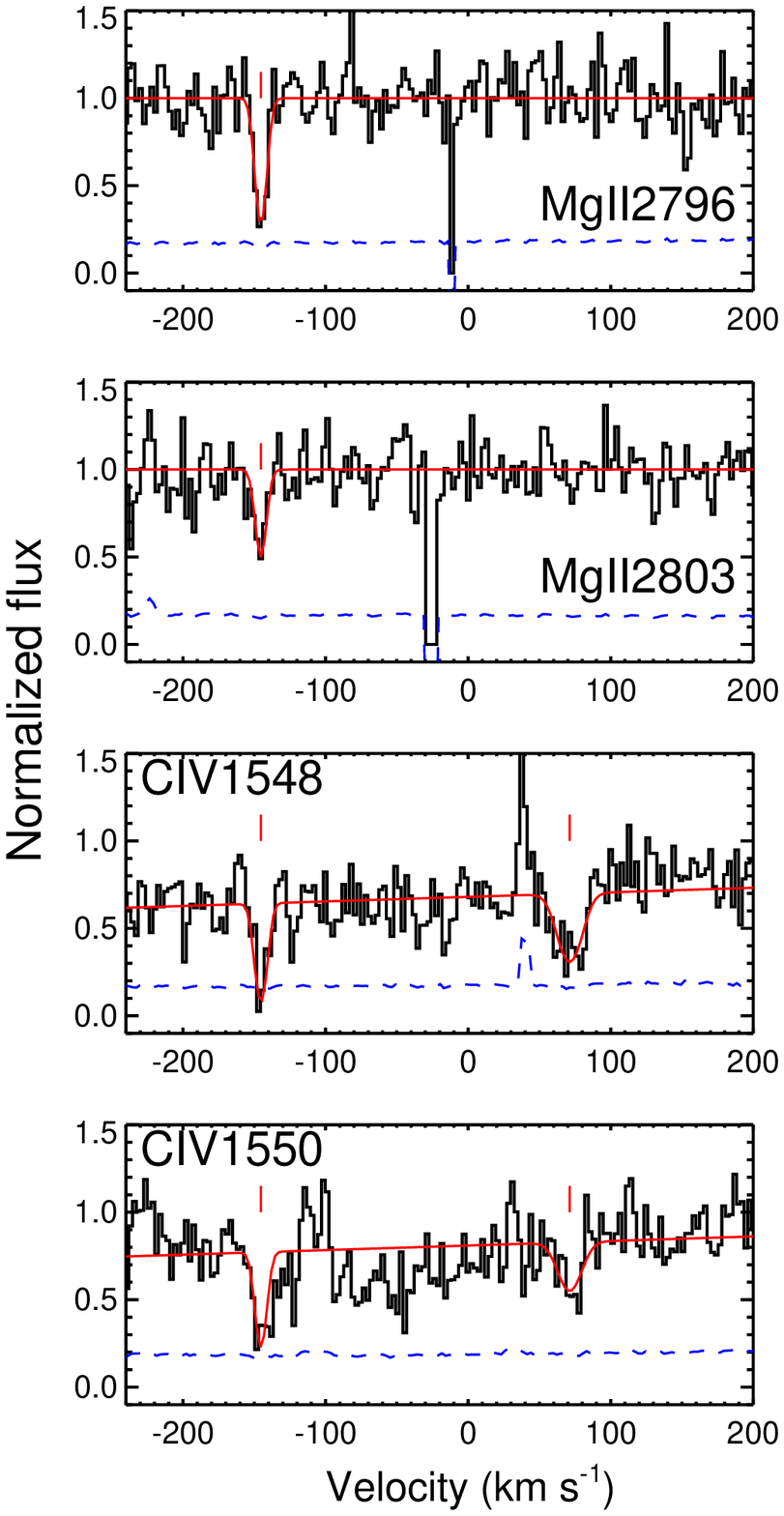}
	\caption{Keck/HIRES spectrum of the quasar close to the metal absorption lines of the system at $z\sim1.104$. The zero velocity is set at $z=1.10497$ that is the redshift of the closest galaxy to the sightline at $b=243$\,kpc. The \CIV\ absorption profiles fall over the wing of a DLA at $z\sim1.68$ where we have to modify the continuum by adding a line. The velocity of the absorbing components are indicated by tick marks. The blue dashed-lines present the error spectra. }
	\label{fig_abs_z1p104}
\end{figure}

The Keck/HIRES spectrum of Q0152$-020$ presents a single component \MgII\ absorbing system at \zabs=1.104 with no detectable associated \FeII\ absorption line. Also at the same redshift we find a \CIV\ absorbing system. The \CIV\ absorption lines consist of two components where one is located at the velocity of the \MgII\ but the second at a $\Delta v=216$\,\kms. These metal absorption lines are shown in Fig. \ref{fig_abs_z1p104}. The histogram and the red continuous lines indicate the observed spectrum and the Voigt profile fits, respectively. 
\subsection{Absorbing system at \zabs=1.211}
\begin{figure}
	\centering	
	\includegraphics[width=1.\hsize,bb=144 92 447 678,clip=,angle=0]{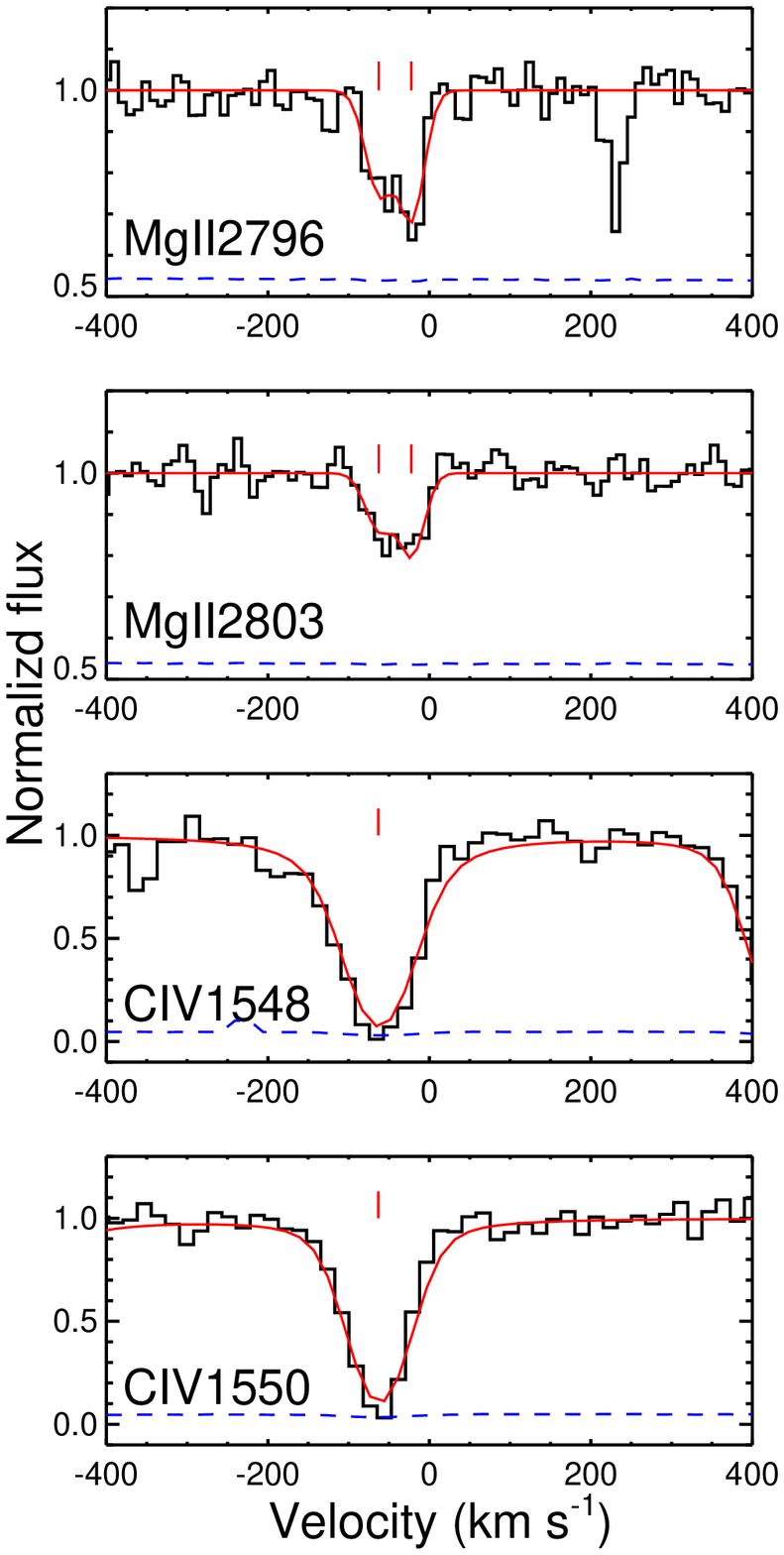}
	\caption{VLT/XShooter spectrum of the quasar close to the metal absorption lines of the system at $z\sim1.211$. The zero velocity is set at $z=1.21134$ that is the redshift of the closest galaxy to the sightline at $b=40$\,kpc. }
	\label{fig_abs_z1p211}
\end{figure}

Neither from the Keck/HIRES spectrum nor from the VLT/X-Shooter do we detect \FeII\ absorption lines associated with this \MgII\ system. However, there exists a strong \CIV\ absorption at the same redshift. Fig. \ref{fig_abs_z1p211} shows the VLT/X-Shooter spectrum of this absorbing system. The best fitting Voigt profile consists of two components for \MgII\ and a single one for \CIV. In passing we note that the higher resolution \CIV\ profile from Keck/HIRES spectrum also does not present any velocity structure as it is fully saturated. 
\subsection{Absorbing system at \zabs=1.224}
\begin{figure}
	\centering	
	\includegraphics[width=1\hsize,bb=144 234 447 537,clip=,angle=0]{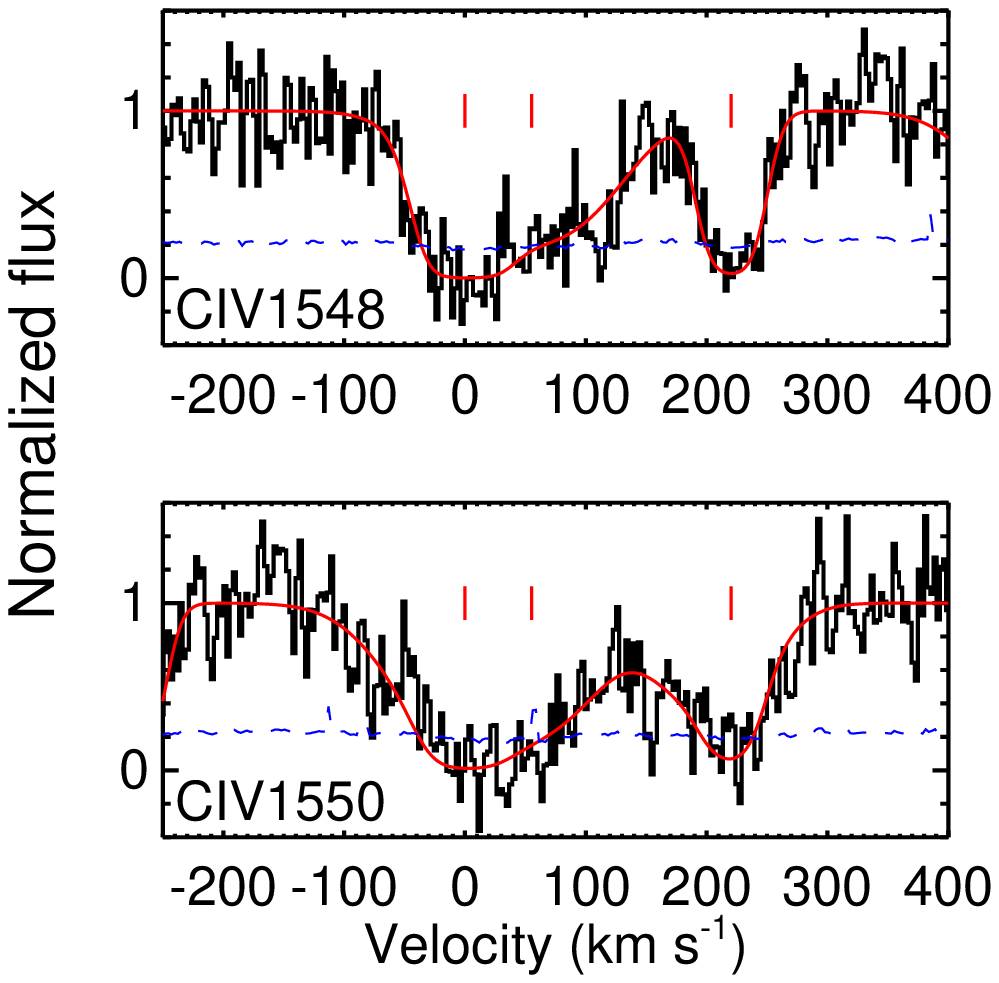}
	\caption{Keck/HIRES spectrum of the quasar close to \CIV\ absorption profiles of the system at $z\sim1.224$. The zero velocity is set at $z=1.22430$. We have not detected any host galaxy close to the redshift of this absorbing system in our MUSE observation.}
	\label{fig_abs_z1p224}
\end{figure}

Keck/HIRES spectrum of Q0152$-$020 presents a strong \CIV\ absorber at \zabs=1.224. Neither in the X-Shooter spectrum nor in the Keck/HIRES spectrum do we detect any of the singly ionized species to be associated with this system. Fig. \ref{fig_abs_z1p224} shows the Keck/HIRES spectrum close to the \CIV\ absorption lines. We modeled the absorption profile using three Voigt profile components spread over 220\,\kms. 
\subsection{Absorbing system at \zabs=1.480}
\begin{figure}
	\centering	
	\includegraphics[width=1.\hsize,bb=150 240 447 537,clip=,angle=0]{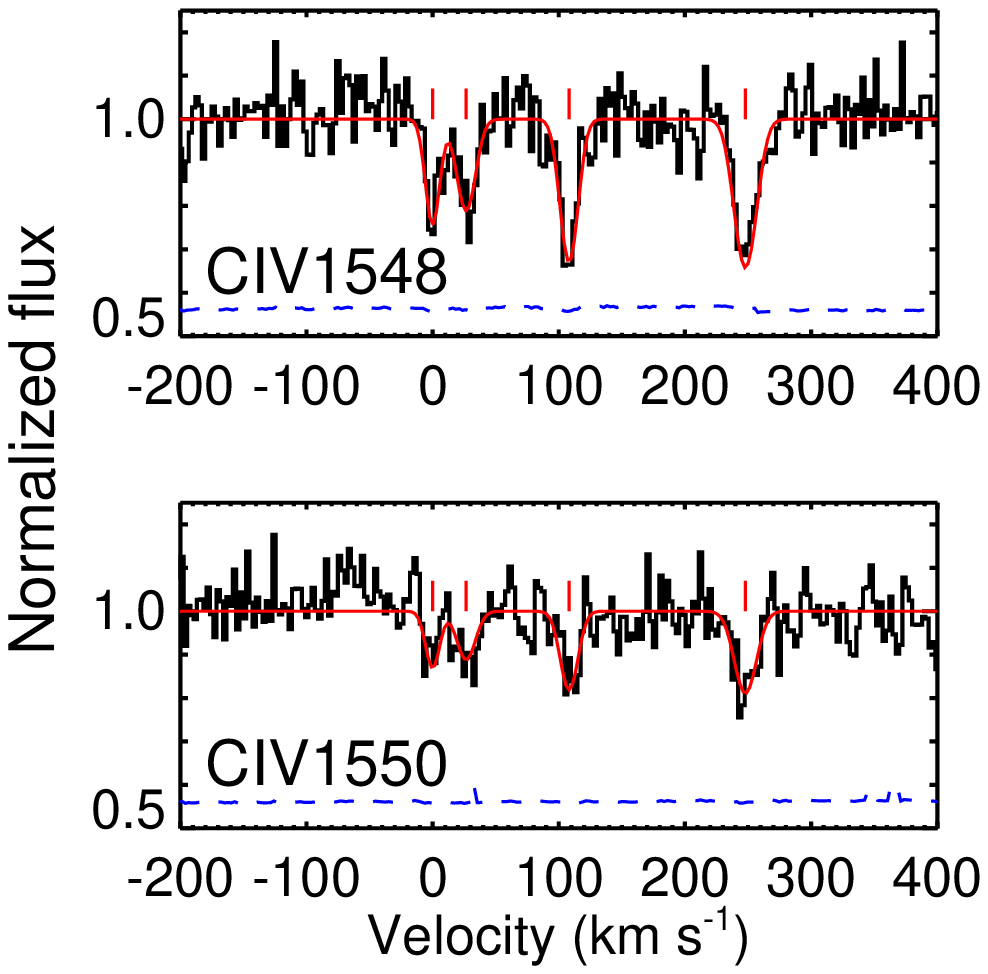}
	\caption{Keck/HIRES spectrum of the quasar close to \CIV\ absorption profiles of the system at $z\sim1.481$. The zero velocity is set at $z=1.47981$. We have not detected any host galaxy close to the redshift of this absorbing system in our MUSE observation.}
	\label{fig_abs_z1p481}
\end{figure}

We do not find any of the singly ionized species to be associated with this relatively weak \CIV\ absorber at \zabs=1.480. The best fitting Voigt profile consists of four components spread over 250 \kms. Fig. \ref{fig_abs_z1p481} shows the Keck/HIRES spectrum of this \CIV\ absorbing system. 
%
%
\section{Host galaxy absorbers}\label{appendix_hostgal}
In this section we present a summary of host galaxy absorbers at $1.0<z<1.5$ towards the sightline of Q0152$-$020. We have only detected [\OII] emission lines from the host galaxies in the cases of the systems at $z=1.104$ and $z=1.211$. With no other detections, we do not have the information required to correct the [\OII] fluxes for internal dust attenuation. Since we do not detect the host galaxy absorbers associated with absorbing systems at $z=1.224$ and $z=1.480$ we report the upper limits for their [\OII] fluxes. Table \ref{tab_other_abs} summarizes the extracted properties of these host galaxy absorbers. We have used the calibration of \citet{Kewley04} to convert the [\OII] flux to the SFR.
\subsection{Host galaxy absorbers at $z\sim1.104$}
\begin{figure}
	\centering	
	\hspace{-1cm}
	\includegraphics[width=1.1\columnwidth,bb=-54 144 666 648,clip=,angle=0]{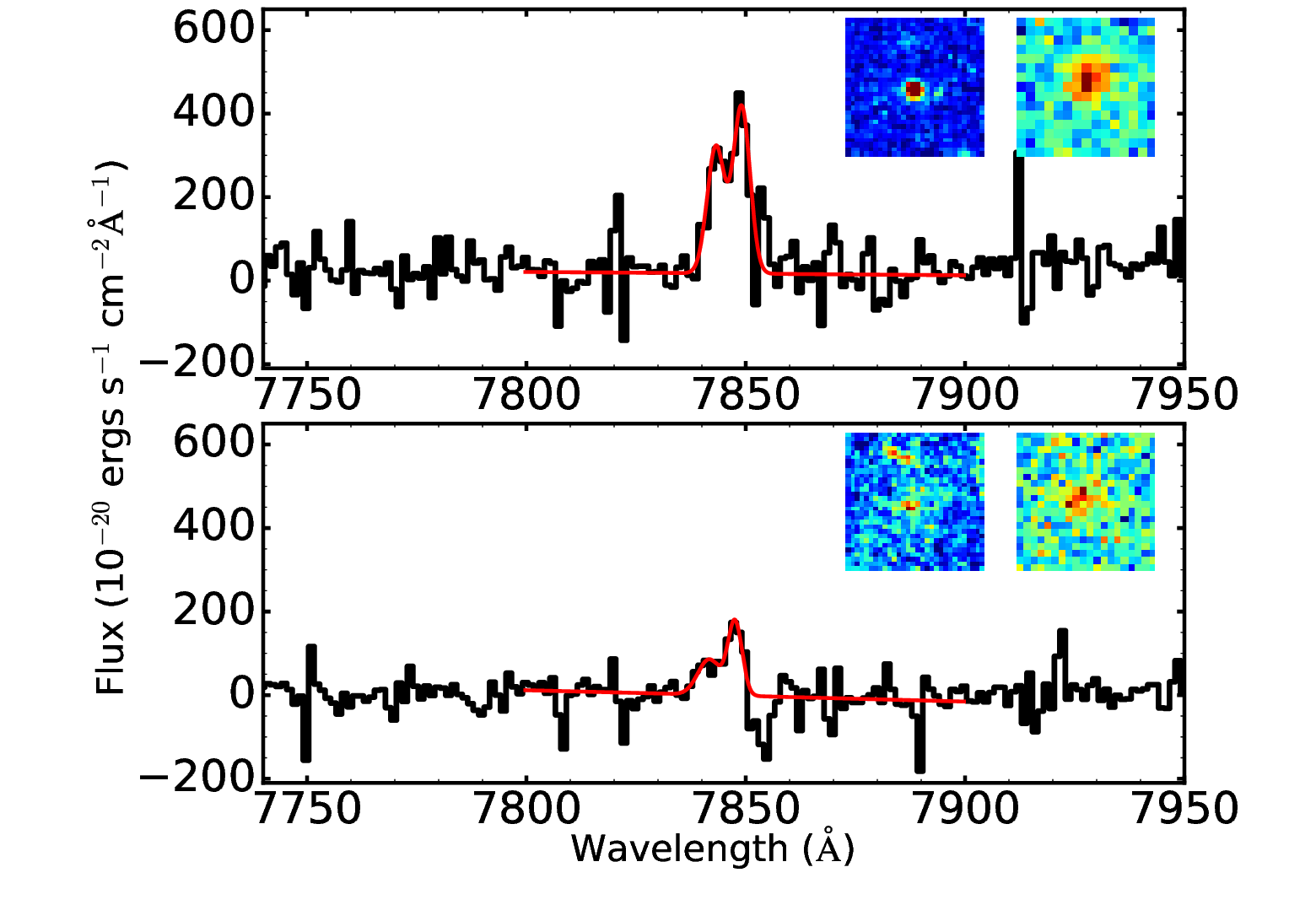}
	\caption{Spectra of the two [\OII] emitters at $z=1.104$. HST broad-band and MUSE [\OII] narrow-band images of the emitters each with a size of $3"\times3"$ are presented on top right in each panel. The histogram and red curves in each panel present the observed and double Gaussian fitted to the [\OII] emission, respectively. On \textit{top-left} side of each panel we present the impact parameter and the velocity separation of the galaxy with respect to the absorber. }
    \begin{picture}(0,0)(0,0)
		\put(-85,265){$b$=243\,kpc, $\Delta v$=145\,\kms}
		\put(-85,183){$b$=261\,kpc, $\Delta v$=97\,\kms}		
    \end{picture}
	\label{fig_oii_z1p104}
\end{figure}

We detected two [\OII] emitters with redshifts close to the absorption redshift. The two galaxies are at large impact parameters of 243\,kpc and 261\,kpc. Fig. \ref{fig_oii_z1p104} presents the [\OII] emission profile from these two emitters. As Fig. \ref{fig_oii_z1p104} shows continuum emission is not detected from either of these two galaxies. 
\begin{figure}
	\centering	
	\hspace{-1cm}
	\includegraphics[width=1.\columnwidth,bb=-54 0 666 792,clip=,angle=0]{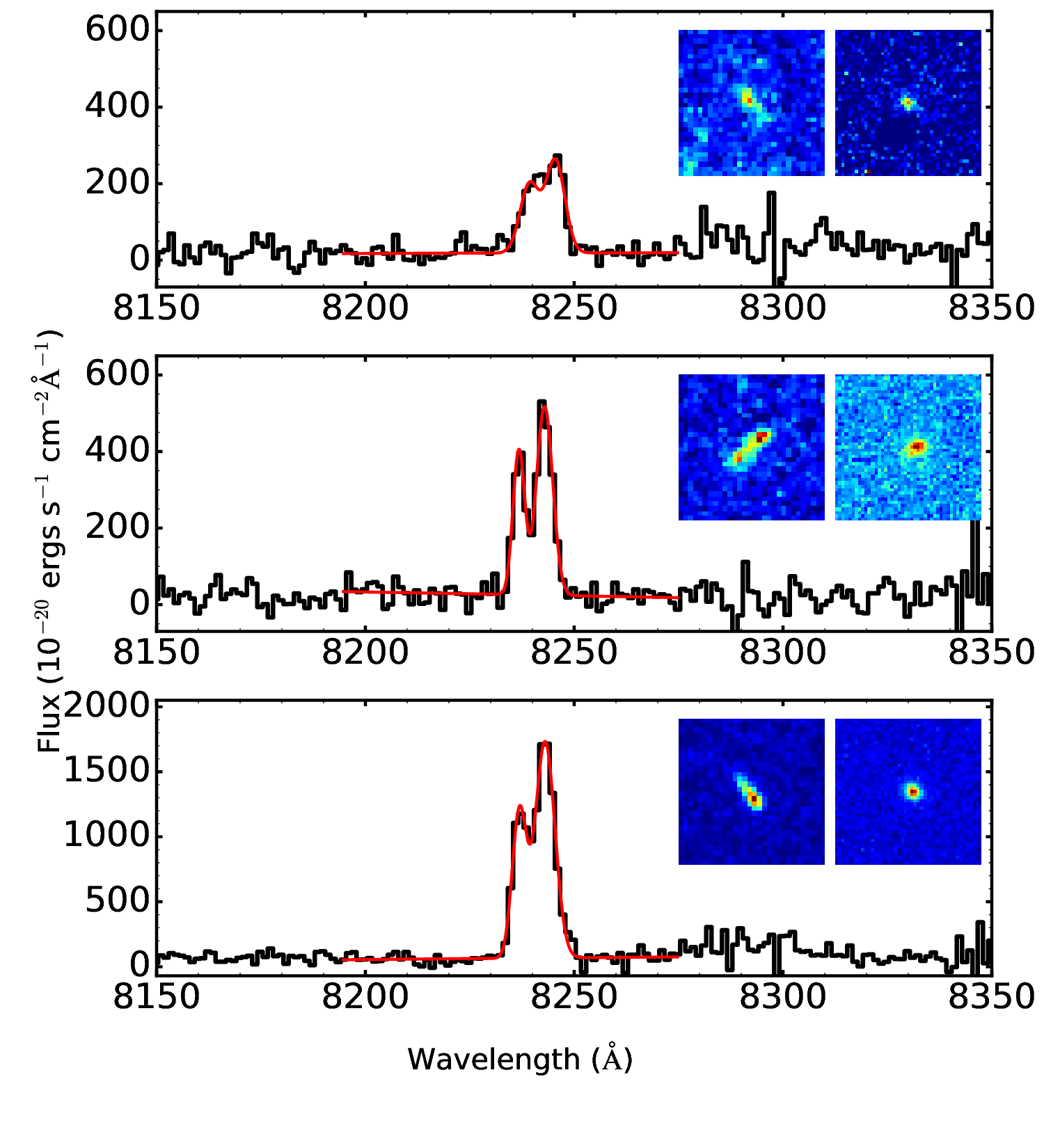}
	\caption{Same as Fig. \ref{fig_oii_z1p104} but for the system at $z=1.211$. 
	}
    \begin{picture}(0,0)(0,0)
	\put(-90,285){$b$=40\,kpc, $\Delta v$=22\,\kms}
	\put(-90,208){$b$=71\,kpc, $\Delta v$=$-$79\,\kms}		
	\put(-90,128){$b$=164\,kpc, $\Delta v$=$-$75\,\kms}		
	\end{picture}	
	\label{fig_oii_z1p211}
\end{figure}
\subsection{Host galaxy absorbers at $z\sim1.211$}
Three [\OII] emitters reside close to the absorption redshifts at impact parameters of 40\,kpc, 71\,kpc and 164\,kpc. Fig. \ref{fig_oii_z1p211} presents the spectra of these galaxies. The galaxy at the largest impact parameter has the strongest emission line and is also detected in the continuum. Inspecting the HST/WFPC2 images of these galaxies we note all the three emitters present perturbed morphologies which can be signatures of recent interactions. In particular for the case of the galaxy at $b=71$\,kpc, as presented in the middle panel of Fig. \ref{fig_oii_z1p211}, we detect signatures of two cores at a distance of $\sim0.6"$ ($\equiv5$\,kpc). 
%
%
%
\begin{table*}
	\small
	\caption{Extracted properties of host galaxy absorbers at $z>1$.}
	\setlength{\tabcolsep}{3pt} 
	\begin{tabular}{lccccccccccccc}
		\hline
		\zabs &$\log$\,$N$(\HI)  &$\log$\,$N$(\MgII) & $\log$\,$N$(\CIV)& ID& RA (J2000)& Dec (J2000)& redshift& $b$  &$\Delta v^1$&  $f$([\OII])$^{2}$ &SFR$^{2}$\\
		&&&&&&&&&$10^{-17}$&\\
		&[cm$^{-2}$]&[cm$^{-2}$]&[cm$^{-2}$]&& hh:mm:ss& dd:mm:ss&&[kpc] &[\kms]& [erg\,s$^{-1}$\,cm$^{-2}$]&[M$_\odot$\,yr$^{-1}$]\\
		\hline
		$1.104$&$<16.8$&$12.48\pm0.08$&$13.74\pm0.09$&G5&01:52:26.3  &	$-$20:01:33  &$1.10497\pm0.00005$ &243& +150&$3.4\pm0.3$ & $1.5\pm0.4$\\
		&&&&G6&01:52:29.1  &   $-$20:01:29  &$1.10463\pm0.00006$  &261&+100& $1.5\pm0.2$ & $0.7\pm0.2$\\\\
		\\
		$1.211$ &$<16.6$&$12.93\pm0.10$&$14.80\pm0.10$&G1&01:52:27.1  &	$-$20:01:02  &$1.21134\pm0.00009$ &40&$+20$& $2.4\pm0.3$ & $1.4\pm0.4$  \\		
		&&& &G3& 01:52:27.5  &	$-$20:00:59  &$1.21059\pm0.00002$ &71&$-80$& $3.7\pm0.1$ & $2.1\pm0.5$  \\		
		&&& &G4& 01:52:25.9  &	$-$20:01:01  &$1.21062\pm0.00002$ &164&$-80$& $14.6\pm0.4$ & $8.4\pm2.1$  \\				
		$1.224$&$<15.5$&$<11.8$&$14.92\pm0.10$ &---&---&---&---&---&---&$<0.4$&$<0.3$\\
		\\
		$1.480$&$<16.0$&$<11.9$&$13.49\pm0.02$ &---&---&---&---&---&---&$<0.3$&$<0.3$\\
		\hline
	\end{tabular}
	\label{tab_other_abs}
	\begin{flushleft}
		$^1$The velocity between the galaxy and the \MgII\ absorption component that has the highest N(\MgII).\\
		$^2$Estimated fluxes; hence the extracted SFRs have not been corrected for dust attenuation in the host galaxies. 
	\end{flushleft}
\end{table*}
\subsection{Host galaxy absorbers at $z\sim1.224$ and $z\sim1.480$}
We do not detect the host galaxies associated with the two \CIV\ systems at \zabs=1.224 and \zabs=1.480. Hence, we estimate the 3$\sigma$ upper limits for the fluxes of the [\OII] emission lines and convert them to the upper limits on the SFR of the possible host galaxies. 

Table \ref{tab_other_abs} presents the flux measurements for different absorbing galaxies. It is worth noting that in the two cases where we do not detect host galaxy absorbers we have not also detected signatures of cold gas (e.g., \MgII\ or \FeII\ lines) to be associated with the absorber. This may imply that such absorbers are produced at impact parameters larger than 250\,kpc that is covered by MUSE or are associated with non star forming galaxies \citep[e.g.,][]{Borthakur13,Bordoloi14_COS}.  
\begin{figure}
	\centering	
	\includegraphics[width=.45\textwidth,bb=0 100 595 842,clip=,angle=270]{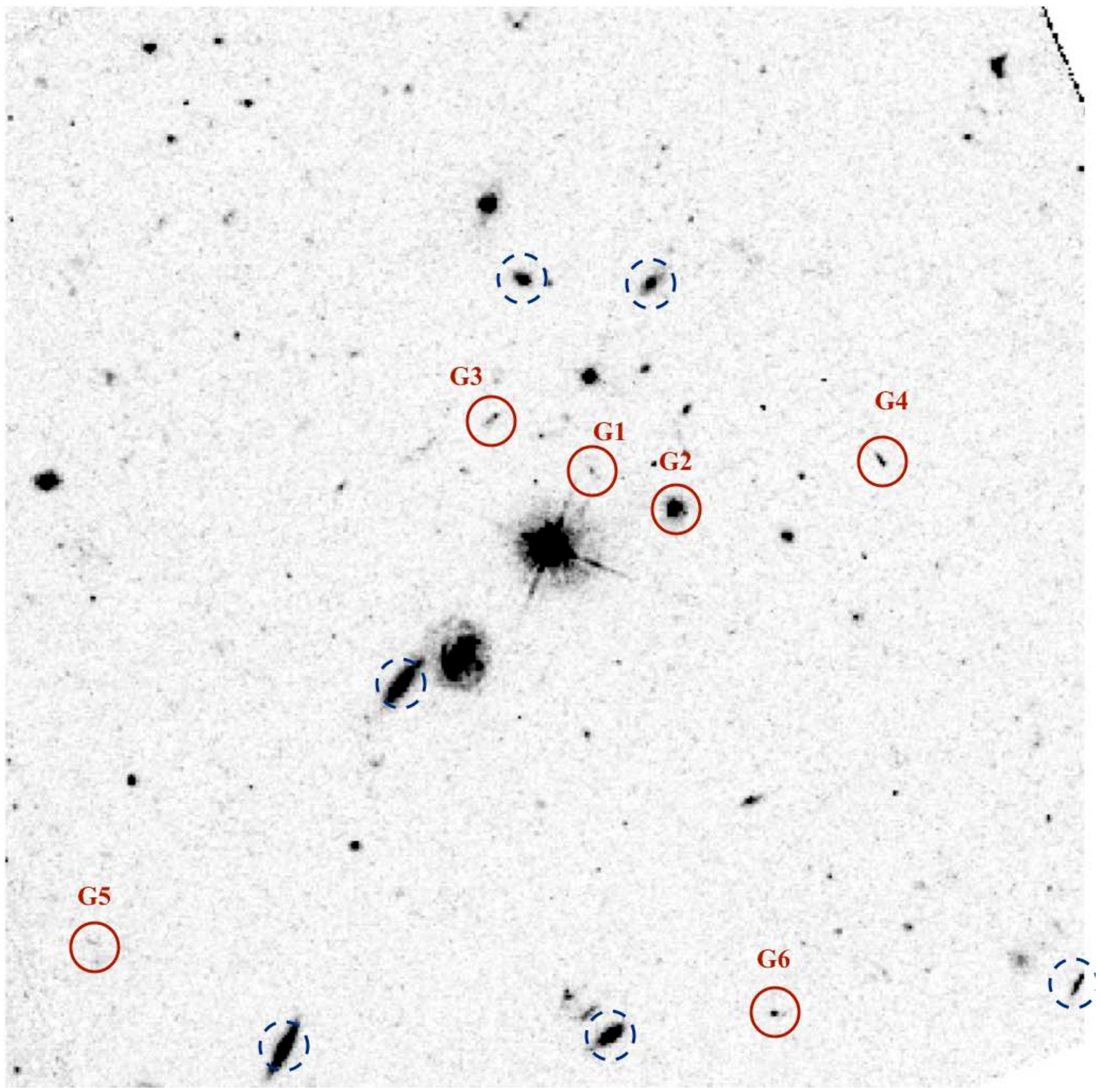}
	\caption{The $1'\times1'$ HST WFPC2 $F720$ image of the field of Q0152$-020$. The ``G1'' to ``G6'' letters mark the sky positions of the galaxies studied in this paper, ordered by impact parameter (see Table \ref{tab_other_abs}). ``G2'' is associated with the absorber at \zabs=0.780. The quasar resides in the centre of the field as indicated by \textbf{Q}. Dashed circles show the positions of galaxies with redshifts close to that of an LLS at $z=0.38$ studied in \citet{Rahmani18}}
	\begin{picture}(0,0)(0,0)
	\put(-3,202){ \textcolor{red}{ \bf\large  Q}} 
	\end{picture}	
	\label{fig_wli}
\end{figure}
\bsp	
\label{lastpage}
\end{document}